\documentclass[journal=jpcafh,manuscript=article,layout=twocolumn]{achemso}
\usepackage[utf8]{inputenc}
\usepackage{esvect}
\usepackage{placeins}
\usepackage[font=footnotesize]{caption}
\usepackage{booktabs}
 
   \usepackage{lmodern} 
   \usepackage{amsmath}   %
   \usepackage{bm}        
   \usepackage{amsfonts}  
   \usepackage{amssymb}   %
   \usepackage{color}
   \usepackage{graphicx}  
   \usepackage{epstopdf}  
   \usepackage{braket}
   \DeclareGraphicsExtensions{.eps}
    \usepackage{multirow} 
    \usepackage{enumitem}

\author{Chiara Panosetti}
\affiliation{Department of Chemistry, Technische Universit{\"a}t M{\"u}nchen, Lichtenbergstr. 4, 85748 Garching b. M{\"u}nchen, Germany}
\altaffiliation{Contributed equally to this work}
\email{chiara.panosetti@ch.tum.de}

\author{Simon B. Anni\'es}
\affiliation{Department of Chemistry, Technische Universit{\"a}t M{\"u}nchen, Lichtenbergstr. 4, 85748 Garching b. M{\"u}nchen, Germany}
\altaffiliation{Contributed equally to this work}
\author{Cristina Grosu}
\affiliation{Department of Chemistry, Technische Universit{\"a}t M{\"u}nchen, Lichtenbergstr. 4, 85748 Garching b. M{\"u}nchen, Germany}
\alsoaffiliation{Institute of Energy and Climate Research (IEK-9), Forschungszentrum J{\"u}lich, 52425 J{\"u}lich, Germany}
\author{Stefan Seidlmayer}
\affiliation{Heinz Maier-Leibnitz Zentrum (MLZ), Technische Universit{\"a}t M{\"u}nchen, Lichtenbergstr. 1, 85748 Garching  b. M{\"u}nchen, Germany}
\author{Christoph Scheurer}
\affiliation{Department of Chemistry, Technische Universit{\"a}t M{\"u}nchen, Lichtenbergstr. 4, 85748 Garching b. M{\"u}nchen, Germany}

\title[DFTB modelling of lithium intercalated graphite with machine-learned repulsive potential]{DFTB modelling of lithium intercalated graphite with machine-learned repulsive potential}

\keywords{\textcolor{black}{battery materials, lithium-graphite, dftb, dftb parametrization, gaussian process regression, machine learning}}

\begin{document}

\begin{tocentry}
\centering
\includegraphics[width=\textwidth]{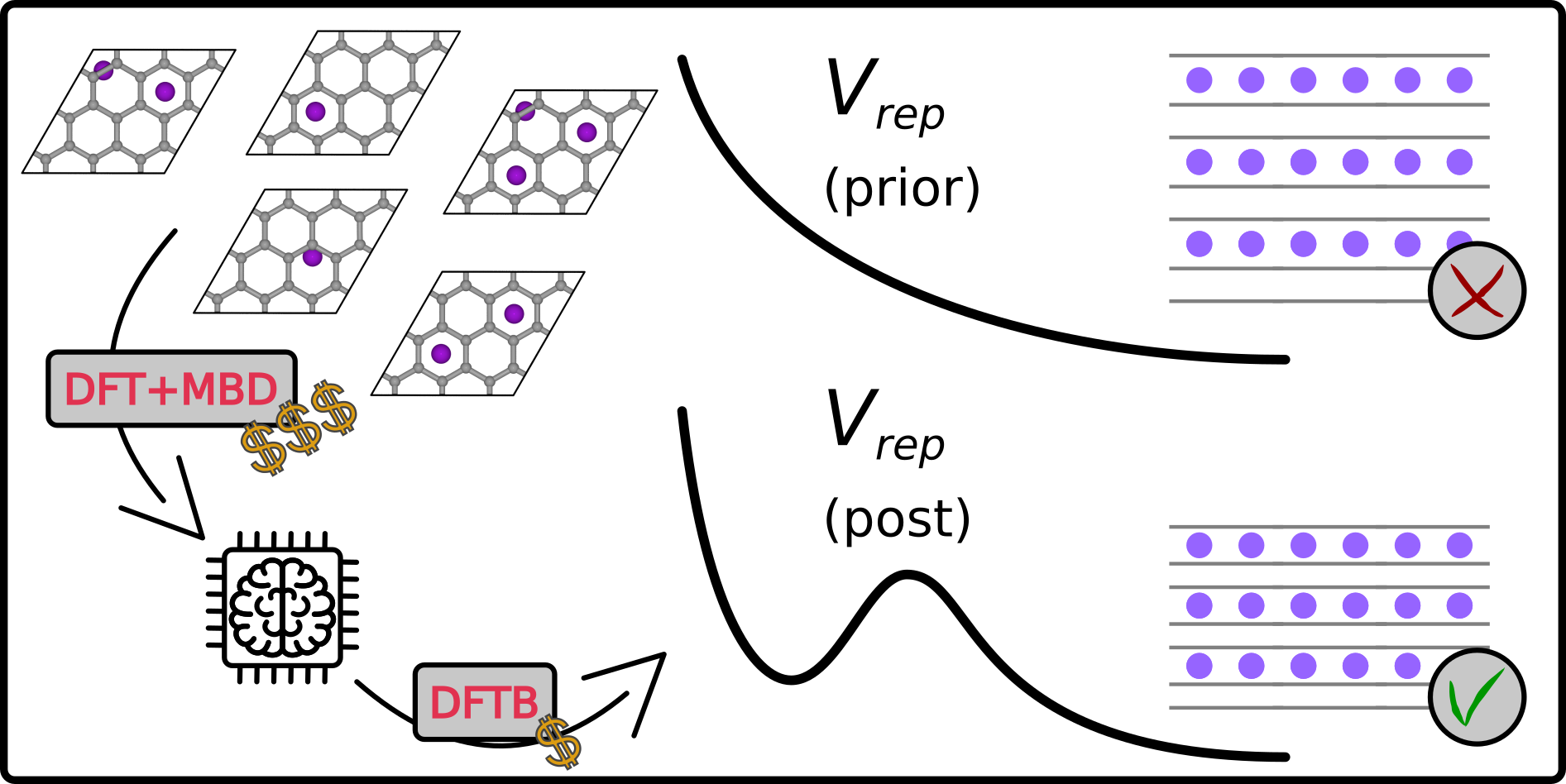}
\end{tocentry}

\begin{abstract}
 Lithium ion batteries have been a central part of consumer electronics for decades. More recently, they have also become critical components in the quickly arising technological fields of electric mobility and intermittent renewable energy storage. However, many fundamental principles and mechanisms are not yet understood to a sufficient extent to fully realize the potential of the incorporated materials. The vast majority of concurrent lithium ion batteries make use of graphite anodes. Their working principle is based on intercalation---the embedding and ordering of (lithium-) ions in the two-dimensional spaces between the graphene sheets. This important process---it yields the upper bound to a battery's charging speed and plays a decisive role for its longevity---is characterized by multiple phase transitions, ordered and disordered domains, as well as non-equilibrium phenomena, and therefore quite complex. In this work, we provide a simulation framework for the purpose of better understanding lithium intercalated graphite and its behaviour during use in a battery. In order to address the large systems sizes and long time scales required to investigate said effects, we identify the highly efficient, but semi-empirical Density Funtional Tight Binding (DFTB) as a suitable approach and combine particle swarm optimization (PSO) with the machine learning (ML) procedure Gaussian Process Regression (GPR) to obtain the necessary parameters. Using the resulting parametrization, we are able to reproduce experimental reference structures at a level of accuracy which is in no way inferior to much more costly \emph{ab initio} methods. We finally present structural properties and diffusion barriers for some exemplary system states.\\
\end{abstract}

\section{Introduction}
\label{intro} 

Within the past decade, studies investigating the consequences of man-made climate change~\cite{Sharp2011, Fisher2012, Program2018} have become more specific, the predicted time frames shorter and the warnings more urgent. The immediate and radical reduction of carbon dioxide emissions by replacing fossil fuel based energy sources with renewable ones has been found to be the only reasonable approach to at least limit those consequences.~\cite{Anderson2016} While the generation of electric energy from wind and sun is already quite advanced and efficient, its storage and transport are the main factors holding it back compared to coal and oil. Currently, two main approaches are being pursued in order to eliminate these drawbacks. One aims directly at the synthesis of alternative liquid or gas-phase fuels. The other intends to improve upon existing battery technology---especially lithium ion batteries---enough, to make it a serious contender in terms of energy sustenance. In this work, we intend to lay some groundwork for gaining deeper insight into some of the atomistic mechanisms limiting the (dis-)charging speed and lifetime of the most common types of lithium ion batteries, with graphite intercalation anodes. 

\begin{figure}[t]
\centering
\includegraphics[width=0.95\linewidth]{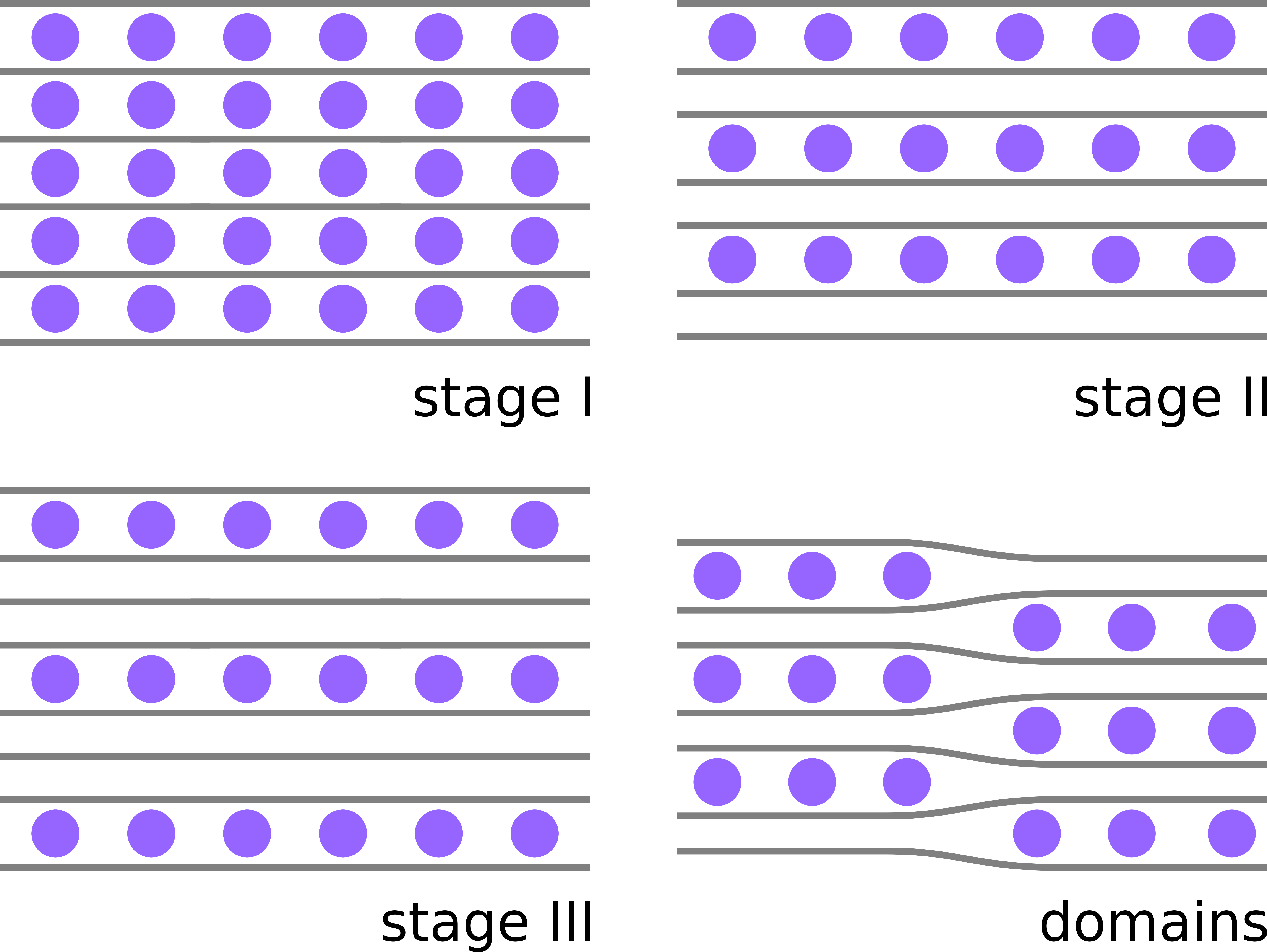}\\
\caption{Sketch of Li-intercalated graphite in stage I to III configurations~\cite{Smith2017}. Violet spheres represent lithium ions, black lines correspond to graphene sheets. Bottom right: illustration of the domain model~\cite{Daumas1969}. The structure has the same nominal stoichiometry as the structure in stage II (top right). 
\label{fig:staging}}
\end{figure}

Ever since graphite was ascertained experimentally and theoretically to be an excellent candidate as an anode for Li-ion batteries, numerous attempts were made at fully describing the working system.~\cite{Hennig1959, Guerard1975,Hawrylak1984,Conard1994,Nitta2015} Most of the electrochemical properties of the anode material itself are well-known. However, in particular transport processes during strongly driven operating conditions, like fast charging, are only poorly understood at a microscopic level. These technologically important macroscopic conditions are accompanied {\em e.g.} by temperature variations, leading to a capacity fade during ageing, as well as lithium plating. All of the above limit the lifetime of the battery.~\cite{Gallagher2016, Wandt2018, Yang2018} Against this background, experiments and theory are pushed quite far to gain insight into the real processes occuring during the electrochemical operation. Depending on the quantities accessible via experiments and theory, two hypotheses are regularly invoked to explain the findings in the range of 0\,\% (graphite) to 100\,\% (LiC$_6$) state of charge~(SOC): the staging and the domain model. The lithium intercalation process shows evidence of multiple phase transitions in the voltage vs.\ SOC diagram. The corresponding system configurations are termed ``stages'' I, II and so forth. In the simple staging model, these correspond directly to the numbers of empty galleries (spaces between graphene sheets) between the fully occupied ones (see Figure~\ref{fig:staging}).%
In the domain model, these motifs are not assumed to range over meso-/macroscopic dimensions but to form regions of finite lateral extent.
Consequently, it is quite clear that different SOC with the same nominal stoichiometry LiC$_x$ will not be configurationally homogeneous, making Li-intercalated graphite a profoundly non-trivial system to address.

In order to effectively connect to experimental studies, a theoretical framework for simulating large-scale and long-duration non-equilibrium processes in the graphite anode, based on kinetic Monte Carlo (kMC)~\cite{Andersen2019} simulations is required. The first step towards this goal is gaining the ability to quickly and accurately calculate diffusion barriers on the fly, which is the primary motivation of this work. This requires the ability to reproduce reliably and accurately the layer distances (ideally of all possible configurations, but predominantly of the dilute, low-saturation stages) and the forces affecting the lithium-ions, while the strains within the graphene layers are of lesser importance.

Large-scale atomistic simulations typically pursue force field approaches~\cite{Duin2001} for those systems where energetics and kinetics are well described within the upper end of the SOC range. However, those approaches are limited when it comes to the entire range of different SOC, from extremely diluted stages to fully concentrated ones. Recently, a Gaussian Approximation Potential~(GAP) was reported to be able to describe amorphous carbon well.~\cite{Deringer2017} However, when the latter was later extended to model lithium intercalation,~\cite{Fujikake2018} it became apparent that the insertion of lithium into those host structures requires a non-trivial description of the electrostatic interaction. Contrary to most approaches, including the one presented in this work, Fujikake \emph{et al.} did not treat the full Li-C system, but attempted to model the energy and force differences arising from lithium intercalation separately, and then added them to the carbon GAP. More specifically, their machine learning (ML) process is based on fitting the energy and force differences between identical carbon host structures, but with and without an intercalated lithium atom. However due to the fact that the lithium intercalation energies are significantly larger in magnitude than the electrostatic lithium-lithium interaction energies, they were not able to recover the latter from the data to a satisfactory degree and had to manually add an extra correction term (fitted to DFT) in order to account for those contributions. To avoid similar shortcomings, we rather base our approach on Density Functional Tight Binding (DFTB)~\cite{Elstner1998}, a semi-empirical---and thus computationally much cheaper---approximation to Density Functional Theory (DFT),~\cite{Kohn1965} which has been the most common technique for high-accuracy electrochemical simulations for many decades~\cite{Koskinen2009}. However, since DFTB's speedup is achieved by pre-calculating atomic interactions to avoid expensive integrations at runtime, this comes at the cost---or rather, initial investment---of pairwise parametrization. As of now, no Li-Li and Li-C DFTB parameters are available. In the following, we combine for the first time the recently developed Particle Swarm Optimization (PSO)~\cite{Shi1998} parametrization approach as first proposed by Chou {\em et al.}~\cite{Chou2016} with a more flexible ML repulsive potential~\cite{Engelmann2018}, to obtain finely-tuned parameters for this system---taking advantage of its physics, albeit perhaps at the expense of some transferability. Let us however stress that the parametrization procedure employed here remains completely general, as the system specificity lies entirely in the choice of the training set(s).

\section{Methods}

\subsection{DFTB: electronic part}
\label{electro}

\begin{figure*}[t]
\centering
\includegraphics[width=0.95\linewidth]{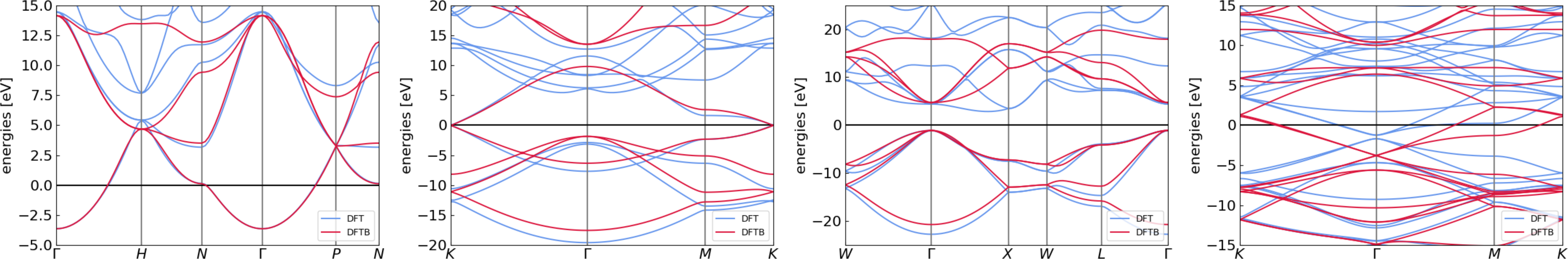}
\caption{Comparison of band structures calculated with PBE-DFT (blue) and our DFTB electronic part parameters (red) for metallic lithium, graphene, diamond and LiC$_6$ (left to right). The latter was not part of our cost function and serves as validation. All the band structures are shifted to the respective DFT Fermi levels.
}
\label{fig:bandstructures}
\end{figure*}

In DFTB jargon, the so-called ``electronic part'' includes the semi-empirical band structure and the Coulombic contributions to the total energy of the system.~\cite{Koskinen2009} These depend parametrically on the diagonal elements $\epsilon$ of the non-interacting Hamiltonian, the Hubbard-$U$ and a confinement potential which is used to cut off the diffuse tails of the basis orbitals. For the free atom, the first two quantities are tabulated for most elements or can be calculated with DFT. However, using the free atom values is an approximation, and the decision whether it is justified must be made carefully on a case to case basis. The confinement potential, on the other hand, is always treated as a parameter. Quadratic~\cite{Stohr2018} and general power-law functional forms~\cite{Wahiduzzaman2013}  are commonly used, as well as the Woods-Saxon potential~\cite{Chou2016} (also employed here) which assures a smoother transition to zero in the orbital tails. Each of these parameters needs to be determined for every chemical species present in the system of interest, typically in a non-linear optimization process. In the PSO, each particle then represents a set of parameters ($\left\{\epsilon\right\}$, $\left\{U\right\}$, and the confinement constants), with which the DFTB interaction is constructed, so that the parametrization can be improved by minimizing a cost function. The central task is thus the definition of a meaningful cost function. Frequently, one uses the weighted sum of an arbitrary number of contributions $f(\sigma^{DFT}, \sigma^{DFTB})$, each providing a measure of the deviation between DFT and DFTB for some system property $\sigma$. Hereby, as we are optimizing the electronic parameters only, the chosen target properties must not depend on repulsion. For our system, we target the band structures of metallic lithium, graphene and diamond. Additional details on the definition of the corresponding cost function, as well as the resulting optimal values of the onsite energies $\epsilon$ and the confinement coefficients, are provided in the SI. Figure~\ref{fig:bandstructures} shows our resulting band structures. Overall, we recognize decent agreement for all band structures, while some deviations are expected given the minimal basis in DFTB. For example, the pronounced mismatch in the conduction band at the $H$ point in the lithium band structure as well as the incorrectly direct band gap of diamond can be ascribed to this over-simplification in the DFTB model. For the two carbon systems, we see very good qualitative agreement for most regions of the band structures, but notice a small degree of overall compression towards the Fermi level. 

Given the overall agreement and also considering the fact that the repulsion potential is capable of quite effectively correcting small imperfections in the electronic part, we decide not to optimize the latter any further in this work---a decision justified in retrospect by the excellent results we present. However, let us still emphasize the opportunity for improvement here, should it eventually become necessary.

From a more techincal standpoint, we note in passing that while during the PSO optimization we employed an $\{sp\}$ basis set for lithium, the production Slater Koster table was constructed including only the $s$ orbital for lithium (with the confinement optimized in the $\{sp\}$ basis). While this may strike as a rather unhortodox choice, it is motivated by the concomitant observation that {\em i)} optimizing the lithium confinement with the $s$ orbital only produces inherently wrong results, and {\em ii)} the optimization of the repulsive potential on top of an $\{sp\}$-basis electronic part showed inherent pitfalls that likely cannot be overcome by any choice of training set. A detailed justification is provided in the SI. 

\subsection{DFTB: repulsive potential}
\label{rep}

It is common practice to assume some analytical form for the repulsive potential and fit the functional parameters as to minimize a set of DFT-DFTB force differences~\cite{Koskinen2009}---a protocol easily implemented also for the PSO approach. However, limitations and bias may result from the choice of said parametrized functional form. It needs to be sufficiently flexible to cover a large space of systems and bonding situations. This typically yields a high dimensional non-linear optimization problem, which might still be insufficient to capture unexpected subtle, yet extremely relevant physical features. We rather adopt the {\tt GPrep} method recently developed in our group~\cite{Panosetti2020, Engelmann2018}, which employs Gaussian Process Regression (GPR)~\cite{Rasmussen2006} to create a flexible functional form ``on the fly'', while adapting to the physics captured by the training data set, instead of forcing us to guess it \emph{a priori}. In the SI, we give a short introduction to the method and explain the character and effect of the related hyperparameters, referring the reader to Rasmussen~\cite{Rasmussen2006} for the underlying stochastic theory and to~\cite{Panosetti2020, Engelmann2018} for the application to DFTB repulsive potentials. For the global damping, correlation distance, and data noise hyperparameters, we verified (see SI) that results are appropriately robust in a sizeable subspace of the overall hyperparameter-space. The same is not necessarily true for the cutoff radii $R_{cut}$. Since the electronic energy contribution is entirely based on just a sum of non-interacting atomic contributions, the repulsion potential has to account for different chemical environments affecting the same type of atom. In a GPR setting it is therefore of paramount importance to sample a sufficiently large set of training data which covers all interatomic distance ranges and chemical environments relevant for a faithful representation of the system studied. Ideally, it should also be ascertained that the model quality is stable w.r.t. the explicit choice of the cutoff radii as well as the other hyperparameters, at least within physically motivated boundaries roughly defined by characteristic lengths of the system, {\em e.g.} nearest neighbour (NN) distances. For instance, it is generally accepted that adequate $V_{rep}$ cutoff values should fall somehwere between the 1st and the 2nd NN distances for the pair in consideration~\cite{Koskinen2009}. However $V_{rep}$ may extend to include ranges beyond the 2nd NN distance, should the particular physics the parametrization is aimed at {\em not} be entirely captured by shorter-ranged repulsive potentials.

\section{Results and discussion}

\subsection{DFTB repulsion training}
\label{training}

In terms of DFT functional, our starting point is PBE~\cite{Perdew1996}, which has been used by the majority of researchers working on intercalation phenomena and is known to describe LiC$_6$ well. However, it does not reproduce the dispersive interaction between graphene sheets. In order to address this, we finally (see ``Set 3'' below) combine the reference PBE calculation with a Many Body Dispersion (MBD@rsSCS, throughout the text referred to as MBD)~\cite{Tkatchenko2012, Ambrosetti2014} treatment and the DFTB model with a computationally cheap Lennard Jones (LJ)~\cite{Zhechkov2005} dispersion correction~\cite{Rappe1992}. The rationale for this choice is that PBE should reproduce galleries containing many lithium atoms correctly and LJ-dispersion should predict empty galleries well, while not interfering too much with the PBE-description of the concentrated ones. However, it is unclear, how this interaction shapes out for intermediate, dilute lithium stoichiometries. During our investigations, we find that this approach works somewhat decently, but needs some controlled adjustments ({\em vide infra}) in order to produce truly satisfactory results.

As a first guess, we construct a set of training structures (Set 1) which consists of a balanced mix of Li$_n$C$_{36}$ super-cells ($n \in (0, 1, ..., 6)$), in order to represent the entire range of charging states. Additionally, those structures are rattled (each atom randomly displaced), as well as compressed or expanded. This procedure yields a smooth distribution of bond lengths and forces. We then train a GPR repulsion potential by matching DFTB against PBE forces for this structural ensemble, aiming at a first, mostly transferable model. The standard LJ DFTB correction is subsequently applied on top of this parametrized DFTB model. With this approach, we are able to find parametrizations that reproduce all layer distances (of graphite, LiC$_{12}$ and of LiC$_6$) correctly, albeit not for a stable range of all parameters.

\begin{figure}[t]
\centering
\includegraphics[width=0.9\linewidth]{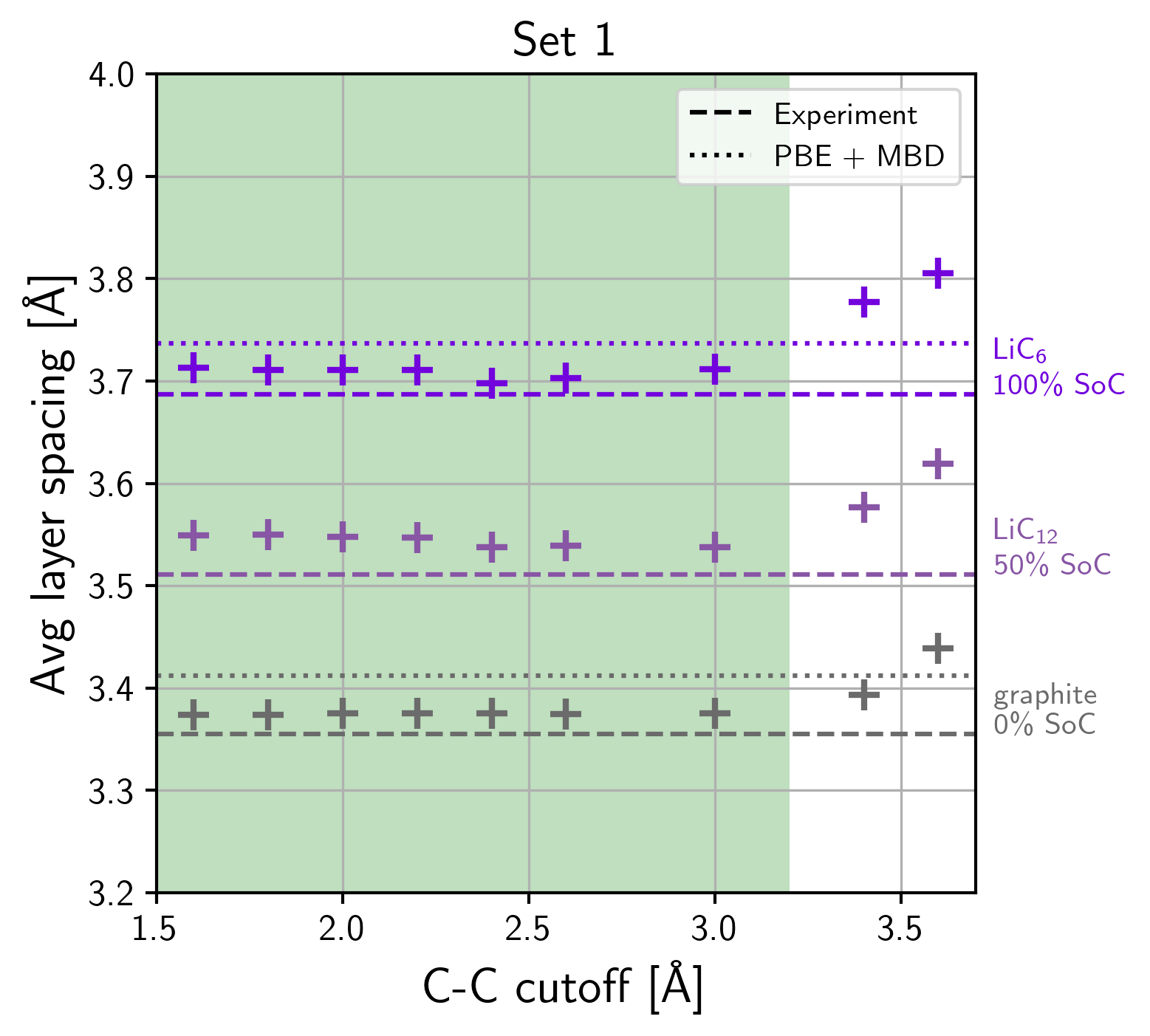}
\caption{Interlayer distances for graphite (grey), LiC$_{12}$ (SOC 50\%, grey-purple) and LiC$_6$ (SOC 100\%, purple) as a function of $R_{cut}^{\mathrm{CC}}$. Note that for LiC$_{12}$ there are two different layer distances to consider: one for the empty gallery and one for the full gallery. Here, we plot the average of the two. The dashed lines show the experimental layer distances we aim to reproduce, as in Trucano {\em et al.}~\cite{Trucano1975} (graphite) and Vadlamani {\em et al.}~\cite{Vadlamani2014} (LiC$_{12}$ and LiC$_6$). The green coloured area represents the range within which the absolute deviation between the DFTB value and the experimental reference is smaller than 0.06~\AA.}
\label{fig:vsCC}
\end{figure}

As shown in Figure~\ref{fig:vsCC}, the choice of cutoff radius $R_{cut}^{\mathrm{CC}}$ for the C-C repulsion potential does not have a major influence on the layer-distances for quite a large range of values. In fact, the point at which the predictions stop being accurate can be identified as approximately the experimental values for the interlayer distances. Going beyond that with the cutoff radius essentially corresponds to including interlayer interactions in the potential fit, mixing their description with the intralayer covalent bonds. Thus, the restriction of the cutoff radius we find here is physically motivated by the range separation of the interactions that characterize our system: as the 2nd next neighbour distance in a relaxed graphene sheet is around $2.45$~\AA{} and the layer distance is $3.35$~\AA{}, the cutoff range defined by the plateau in Figure~\ref{fig:vsCC} represents a sweet spot where the GPR learns 2nd next neighbour interactions but does not yet (mistakenly) take any interlayer interactions (even in the compressed structures) into account in the repulsion potential. However, the same reasoning does not apply to CC bond lengths, which are not correctly reproduced if the GPR learns forces beyond the 1st NN distance (see SI). In light of these findings, we select the cutoff value $2.2$~\AA{} for the C-C-repulsion potential. Indeed, we did not encounter any reason to change this selection during the entirety of this work (despite rigorously testing it for each of the training data sets). 

However, with this first training set we do not obtain an equally stable plateau as a function of the Li-C repulsive cutoff (see SI), with the correct values corresponding to $R_{cut}^{\mathrm{LiC}} = 4.0$~\AA{} not belonging to a plateau at all. Furthermore, the quite strongly distorted graphite planes in these structures lead to large forces compared with those acting on the intercalated lithium-ions, causing a systematic underestimation in lithium-forces prediction. We tackle the second problem first: while the rattled, scaled structures in Set 1 cover a sufficiently large range of bond lengths, they only account for configurations with the lithium-ions sitting over the centre of a graphite ring, \emph{i.e.} in a local energy minimum. We recognize this as the reason for the comparably small lithium-forces. In order to balance out this structural bias, we calculate a number of transition paths for lithium diffusion processes in LiC$_6$ and LiC$_{12}$ stage I/II compounds using a Nudged Elastic Band (NEB) method~\cite{Henkelman2000, Henkelman2000a}. We are now able to extract structures from these trajectories, in which the lithium ions are subject to stronger forces commensurable with the graphite-layers. For our second training set (Set 2), we replace the higher-saturated rattled and scaled structures with those extracted from the transition paths. In doing so, we assume higher-saturated structures to be responsible for the slight contraction observed in C-C bonds (see SI).

\begin{figure}[t]
\centering
\includegraphics[width=0.95\linewidth]{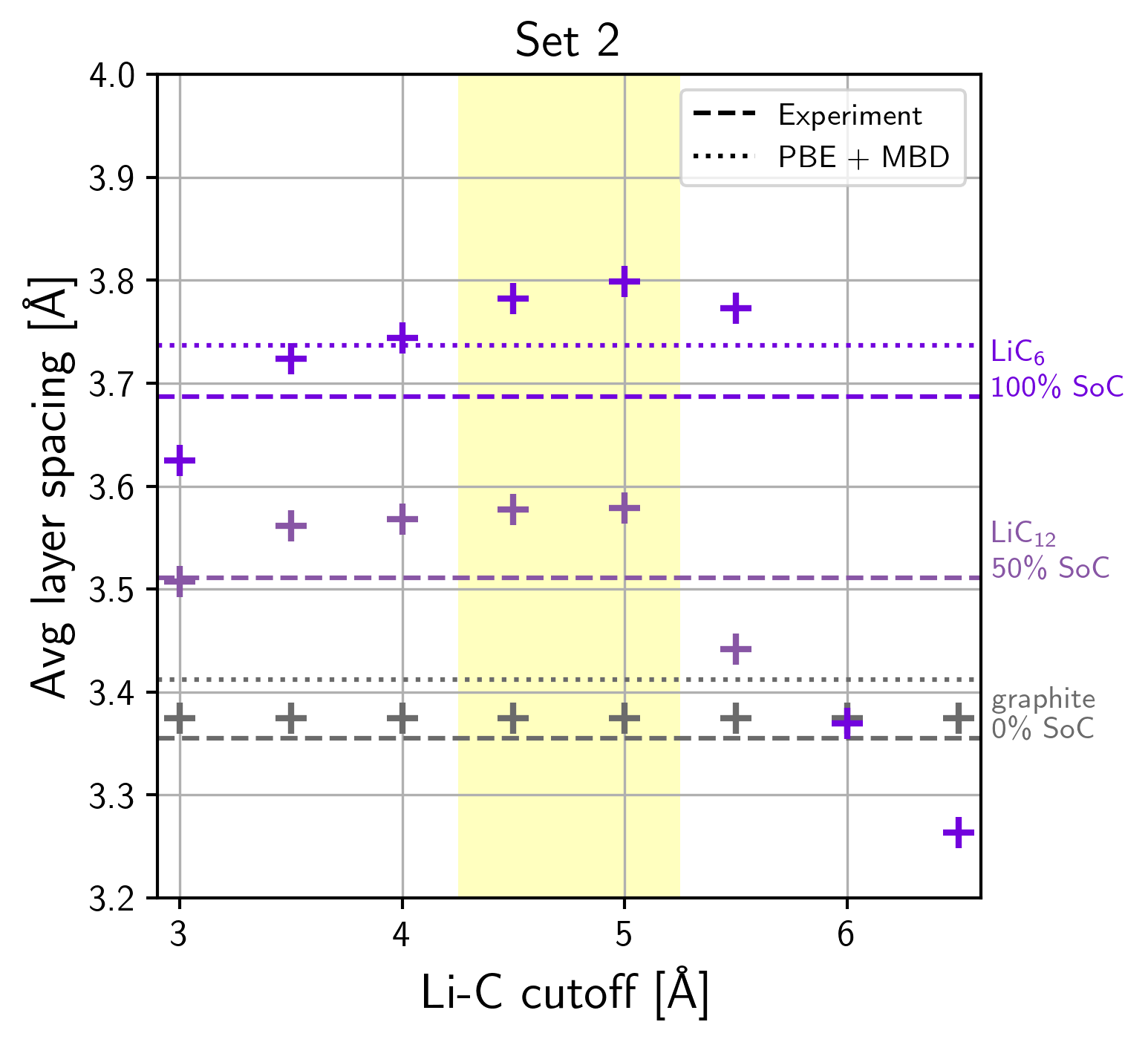}
\caption{Interlayer distances for LiC$_{12}$ (SOC 50\%, grey-purple) and LiC$_{6}$ (SOC 100\%, bright purple) as a function of $R_{cut}^{\mathrm{LiC}}$, with a fixed $R_{cut}^{\mathrm{CC}}$ set to 2.6~\AA{}. The repulsion was trained on a set analogous to Set 1 ({\em cf.} text), where the higher-saturated structures were replaced by geometries randomly extracted from intra-layer Li diffusion paths. For LiC$_{12}$, the plotted interlayer distance is the average between the values for the filled and the empty gallery. The dashed lines show the experimental layer distances. The yellow coloured area represents the range within which the results are stable, however at a wrong value.}
\label{fig:vsLiC}
\end{figure}

By this measure, we are able to improve the accuracy for predicting forces on Li-ions significantly (albeit still slightly underestimated), without sacrificing the description of the graphite layers. However, while we do observe a plateau for the resulting layer distances with respect to $R_{cut}^{\mathrm{LiC}}$, the interlayer distances are not reproduced equally well as in Figure~\ref{fig:vsCC} for Set 1 (see Figure~\ref{fig:vsLiC}, yellow area), with the exception of points 3.5~\AA{} and 4.0~\AA{} which do not belong to a plateau. This behaviour suggests that our problem here does not lie in the choice of the training set, but rather in the treatment of long-ranged interactions.

Let us consider the underlying predicament: so far, the DFTB-part of the force residues used for the ML process is calculated without LJ dispersion correction. We then construct the repulsion potential with the purpose of making those DFTB calculations match references based on PBE-DFT, which reliably predicts layer distances for LiC$_6$. By then using LJ (required to obtain the correct empty layer distance in graphite) in our actual DFTB calculations (after the parametrization process), we cause the aforementioned offset for highly lithiated compounds. Using LJ already for the force-residue calculations during the ML seems like the obvious solution to this problem. However, this presents a new issue in the lower-saturation range (LiC$_x$, $x>12$). There, we previously fitted the repulsion to PBE-DFT references, which are not correct in that range without dispersion correction. The resulting DFTB forces are then shifted by LJ towards the correct value (as is indicated by the quite decent results for LiC$_{12}$ with Set 2). But after the modification, we would then fit the \emph{final} DFTB forces (that result after applying the LJ) to the (incorrect) PBE-DFT references, thus improving our performance for highly saturated system states, but ruining it for dilute ones, by effectively double counting dispersive contributions. It becomes apparent that in order to make this approach work, we need to utilize dispersion corrected DFT reference forces which are also correct for low saturation states and, at the same time, compatible with the computationally cheap DFTB-LJ correction.

\begin{figure}[t]
\centering
\includegraphics[width=0.95\linewidth]{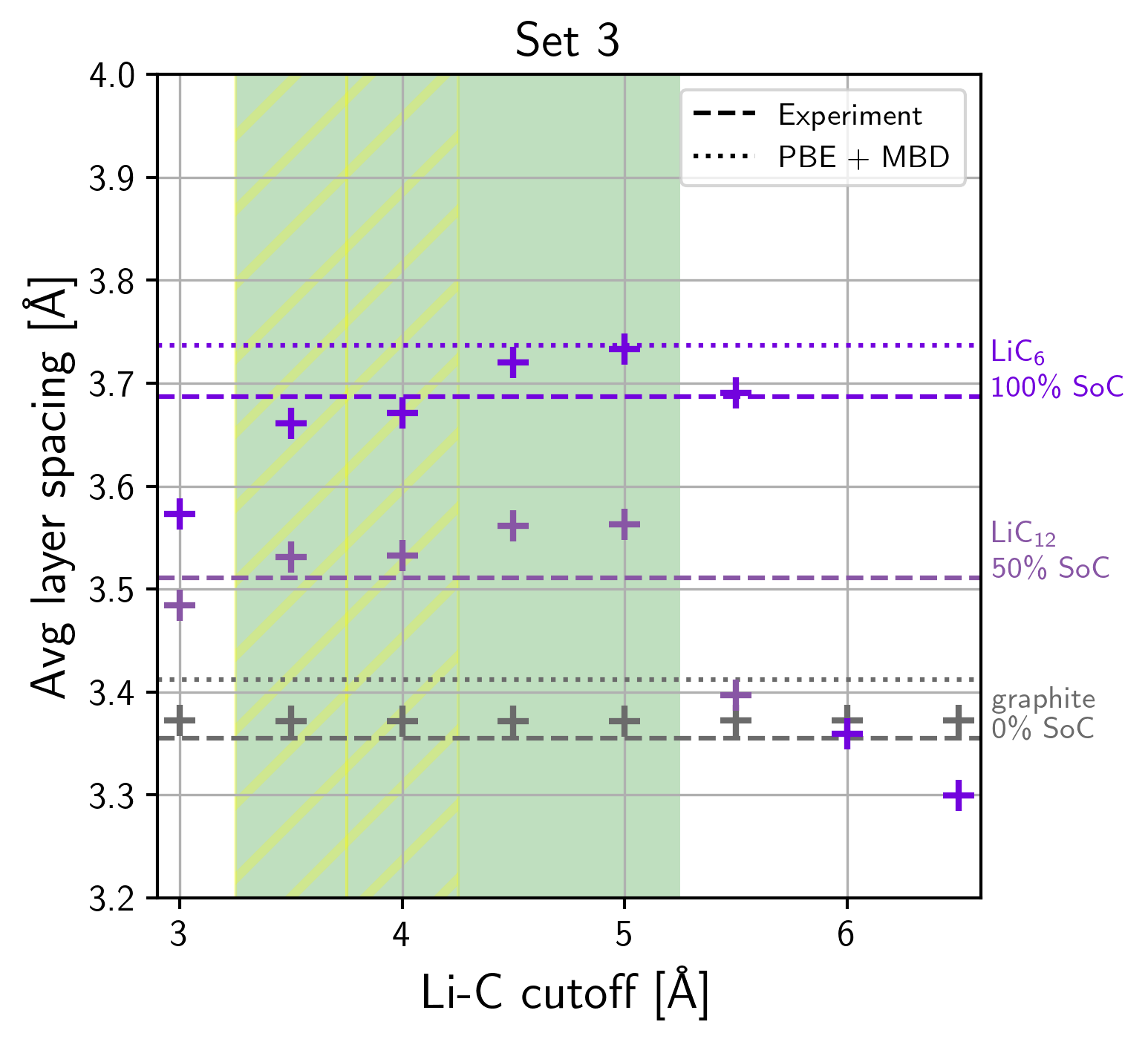}
\caption{Interlayer distances for LiC$_{12}$ (SOC 50\%, grey-purple) and LiC$_{6}$ (SOC 100\%, bright purple) as a function of $R_{cut}^{\mathrm{LiC}}$, with a fixed $R_{cut}^{\mathrm{CC}}$ set to 2.6 \AA{}. The repulsion was trained on a set analogous to Set 2 ({\em cf.} text), where ~70\% of the structures were replaced by geometries with MBD-corrected forces. For LiC$_{12}$, the plotted interlayer distance is the average between the values for the filled and the empty gallery. The dashed lines show the experimental layer distances. The yellow-hatched area represents a range within which the results are stable and correct, however we consider them not ideal.
}
\label{fig:vsLiC_set3}
\end{figure}

\begin{figure*}[t]
\centering
\includegraphics[width=0.95\linewidth]{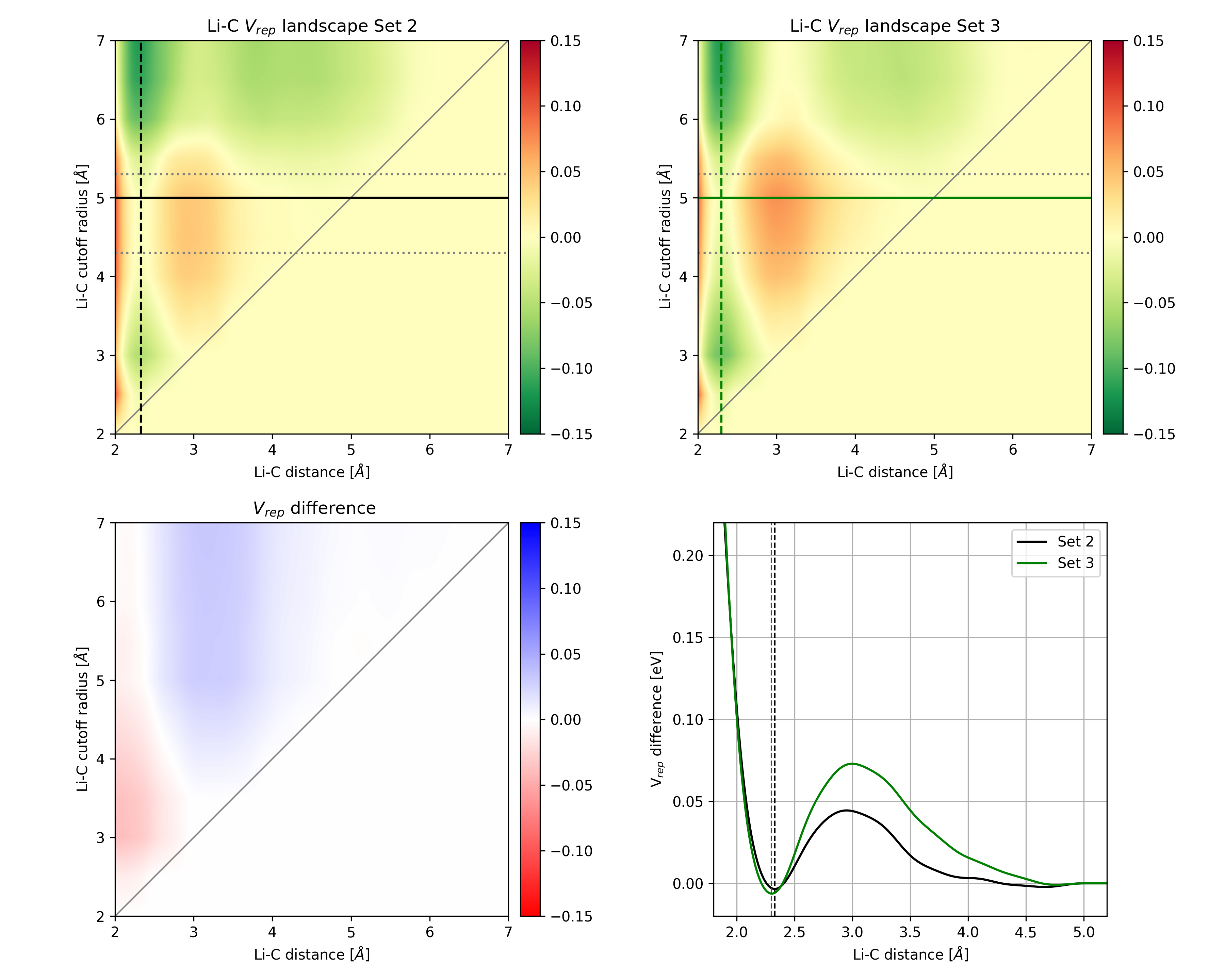}
\caption{Top: 2D repulsion potential landscape (units $eV$) expressed by the colour bar) depending on the chosen Li-C cutoff radius (y-axis) for Set 2 (left) and Set 3 (right). The black and green dashed lines represent the next neighbour Li-C distances for sets 2 and 3 respectively. The diagonal lines illustrate the cutoff radii, at which the potential is set to zero. The plateaus are highlighted between thin dotted lines. Bottom: (Left) Influence of the inclusion of MBD vs LJ force residues in the training data on the repulsion potential (units $eV$ (expressed by the colour bar). (Right) Detailed repulsion potentials at $R_{cut}^{\mathrm{LiC}} = 5.0$~\AA{}.}
\label{2Dlandscape}
\end{figure*}

Our ansatz is that we can---to a degree---encode the difference between the LJ dispersion and the ``true'' dispersion into the repulsion potential. At this point we stress that \emph{ideally}, both the true, non-local exchange correlation functional in DFT and an ideal repulsion energy in DFTB would already encompass all dispersion effects, and it is solely due to approximations in the derivations, {\em e.g.} of GGAs, that they do not in these models. Therefore, rather than mixing our repulsion potential with something fundamentally different (which would be physically questionable), what we do here simply corresponds to partially adding a contribution back in, that should have been there in the first place. To our knowledge, the currently best way to calculate dispersion corrected lithium intercalated graphite, with correct layer distances predicted for the entire saturation range, is the Many Body Dispersion (MBD) correction~\cite{Tkatchenko2012}. This method is computationally rather expensive, but since we only need to run DFT calculations for our training data set, which is very limited in size, this is not vital to us.

In practical terms, we then build a Set 3 where DFTB-DFT force residues are replaced by DFTB(LJ)-DFT(MBD) force residues. We do realize that this approach most likely comes with some cost in terms of transferability. In order to retain as much of it as possible, we choose not to replace \emph{all} force residues, but only $\approx66\%$ (more precisely, only for structures containg no or 1 lithium atom), which proves sufficient to demonstrate the effectiveness of the presented method in a general way. Nonetheless, further investigating the effect this percentage has on the performance is certainly a task that should be tackled in the future. Of course, alternatively to our approach, it is possible to simply apply the MBD correction scheme directly to our DFTB calculations. However, doing so would cost us one to two orders of magnitude in speed, as MBD then becomes the computationally dominating step in production DFTB calculations.

Using the previously explained modifications, we have succeeded at shifting the predicted interlayer distances (within the stable $R_{cut}^{\mathrm{LiC}}$ plateau) into the very close proximity of the experimental reference values for both LiC$_6$ and LiC$_{12}$, as shown in Figure~\ref{fig:vsLiC_set3}. To be precise, the ranges of $R_{cut}^{\mathrm{LiC}}$ = $\sim 3.3-4.3$~\AA{} and $\sim 4.3-5.3$~\AA{} should be regarded as two distinct plateaus, both close to the correct experimental values. However, we shall consider the second as our final plateau, where the resulting interlayer distance fall between the experimental and the DFT reference values, since we trained against DFT reference forces. Of note, the predicted interlayer distance for LiC$_{12}$ becomes wrong already at $R_{cut}^{\mathrm{LiC}} = 5.5$~\AA. Once again this is a physically motivated boundary: it is the distance at which the Li atoms start to ``feel'' the next layer. In LiC$_6$, which has a larger interlayer spacing and for which the resulting interlayer distance is correct at that point, this happens for $R_{cut}^{\mathrm{LiC}} > 5.5$~\AA.

Especially the excellent results for the stage II compound LiC$_{12}$ show that our parametrization is now able to handle \emph{both} mainly ionic concentrated \emph{and} mainly dispersive dilute layers to a satisfactory degree. In Figure~\ref{2Dlandscape}, we illustrate the effect our modification has on the repulsion potential landscape for a wide range of Li-C cutoff radii. First (and most notably), we have moved and solidified the local minimum related to the next-neighbour lithium-carbon interaction (see bottom right). For the Set 2 and Set 3 potentials, the minima (black and green dashed lines respectively) are located at atomic distances of $2.33$~\AA{} and $2.30$~\AA{} respectively, which correspond to LiC$_6$ interlayer distances of $3.80$~\AA{} and $3.73$~\AA{}, the exact values which \emph{do}, in fact, result from the relaxation of those structures, using the two repulsion potentials respectively. The 2D maps (top) show that this behaviour is apparent for an entire range of cutoff radii, thus ruling out the possibility that the fit is only accidentally correct (as it happens, {\em e.g.}, for Set 1 with $R_{cut}^{\mathrm{LiC}} = 4.0$~\AA. We can also clearly see the upper ($\sim 5.3$~\AA) and lower ($\sim 4.3$~\AA) boundaries for the cutoff radius, beyond which the physicality of the model falls apart. They define exactly the range within which we find the stable cutoff dependency plateau, which is now at the correct numerical value as shown in Figure~\ref{fig:vsLiC_set3}. We may identify the upper boundary at $5.3$~\AA{} (as discussed above), as the distance between a lithium ion and the second closest graphene sheet, which is an intuitively plausible limitation. It is less obvious, though, to assign a clear physical meaning to the lower bound at $4.3$~\AA{}, as it cannot be directly related to any particular structural feature of LiC$_x$. The most likely cause, we believe, is that the cosine-shaped cutoff function employed in the GPR framework starts cutting off physically relevant details from the repulsion potential below that.


A physically motivated lower bound of different nature may be identified by evaluating the relative RMSD of forces as a function of Li-C cutoff, shown in the SI.  
Overall, we now observe two separate Li-C cutoff plateaus: between approximately 4.3~\AA{} and 5.3~\AA{}, we obtain accurate layer distances (Figure~\ref{fig:vsLiC_set3}), while for radii above roughly 4.0~\AA{} and until 6.0~\AA{}, our predictions for forces and transition energies are correct. 
This duality can very simply be explained by the fact that the first property is mostly a {\em z}-direction phenomenon (and interactions with the second closest graphene sheet limit the physicality of our model), while the other takes place almost exclusively in the {\em xy}-plane, where no such limitation applies, hence the broader plateau. Given this difference in fundamental nature, it is very plausible to trust both these plateaus. Thus, their overlap (4.3--5.3~\AA{}) defines the region within which any value of the Li-C cutoff radius produces an almost identical parametrization that performs very well, for all our benchmark criteria, in a stable and trustworthy manner.

\subsection{Interlayer distances and diffusion barriers}
\label{results_barriers}

\begin{table*}[t]
\centering
\begin{tabular}{l l | r r r r r r }
\hline
\multicolumn{8}{l}{(Average) inter-layer distances with detailed analysis of layer spacing and barriers.} \\
\hline

&compound	              &experimental  &DFTB          &DFT~\cite{Krishnan2013} &filled &empty &barrier      \\ \hline
graphite		& --          &$3.355$~\AA{} &$+17$~m\AA{}  &$+62$~m\AA{}               &--             &$+17$~m\AA{}  &--           \\
LiC$_{18}$	&stage III    &$3.470^{\dagger}$~\AA{} &$+35$~m\AA{}  &$+173$~m\AA{}              &$+198$*~m\AA{} &$-98$*~m\AA{} &$468$~meV    \\ 
LiC$_{12}$	&stage II     &$3.511$~\AA{} &$+52$~m\AA{}  &$-16$~m\AA{}               &$+148$*~m\AA{} &$-147$*~m\AA{}&$480$~meV    \\
LiC$_6$		&stage I      &$3.687$~\AA{} &$+46$~m\AA{}  &$+56$~m\AA{}               &$+46$~m\AA{}   &--            &$503$~meV    \\ \hline
\multicolumn{8}{l}{\footnotesize{*: note that these numbers are not errors, but differences between specific and average layer distances}}\\
\multicolumn{8}{l}{\footnotesize{${\dagger}$: estimated from experimental values for graphite and LiC$_6$.}}\\
\end{tabular}\\
\caption{Summary of the interlayer distances resulting from our DFTB parametrization (via structure optimization using the BFGS algorithm~\cite{Shanno1985}) compared with experimental values from Trucano {\em et al.}~\cite{Trucano1975} (graphite), Vadlamani {\em et al.}~\cite{Vadlamani2014} (LiC$_{12}$ and LiC$_6$). For the LiC$_{18}$ reference, we assume the filled gallery and the empty galleries to have the same interlayer spacing as LiC$_6$ and graphite respectively. For stage II and III compounds, we consider the average layer distance here. Furthermore, we show the relative deviation of our results and compare them with those reported by Krishnan \emph{et al.}.~\cite{Krishnan2013}}
\label{table1}
\end{table*}

Table~\ref{table1} reports some resulting inter-layer distances and diffusion barriers based on our DFTB parametrization, compared with experimentally determined values, as well as previous theoretical findings.
As a quick reminder, stages I, II and III correspond to every, every other and every third gallery being filled (to any degree) with lithium. Additionally, one may describe the concentration of the intercalant in a filled gallery as dilute (low) or concentrated (high), thus allowing for a simple classification of fundamentally different compounds. Here, however, we take only concentrated stages into consideration.

For all calculations, we chose a Li-C cutoff radius of $5.0$~\AA{}, following the findings discussed above.
As Table~\ref{table1} clearly illustrates, we systematically outperform the method by Krishnan \emph{et al.}~\cite{Krishnan2013}---in terms of accuracy---for every structure they provide comparison for. This is especially remarkable considering the fact that they used full GGA-DFT with dispersion corrections in post-processing, which is the current state-of-the-art approach, as well as significantly more computationally expensive than our method.

Subsequently, we investigate intra-layer next-neighbour diffusion barriers and compare our results to recent experimental findings from Ref.~\cite{Umegaki2017} (based on muon spin relaxation spectroscopy) and theoretical from Ref.~\cite{Toyoura2010} (calculated at the LDA-DFT level without dispersion correction, which is only reliable for the predominantly ionic, filled state of charge).

Our calculations yield purely \emph{microscopic} results within 50 meV from each other for all three relevant compounds, as is shown in Table~\ref{table1}. The deviations between them correlate to the slight differences in the filled-layer spacing of the different structures. Our 503 meV barrier for the elementary diffusion in LiC$_6$ is in perfect agreement with the value of 490 meV reported by Toyoura {\em et al.}~\cite{Toyoura2010} In contrast, the experimentally determined \emph{active} barriers of 270 meV for LiC$_6$ and 170 meV for LiC$_{12}$ show a strong dependency on the systems stage. We believe this difference to be caused by concerted effects. Capturing those using kinetic Monte Carlo simulation is something we intend to do in the near future.

\section{Conclusions and outlook}

In this work, we put forward---for the first time combining particle swarm ({\em i.e.}~PSO) and machine learning~\cite{Engelmann2018} ({\em i.e.}~GPR) approaches for this task---a well-performing DFTB-parametrization for lithium intercalated graphite which is capable of very accurately reproducing various structural properties and qualitative trends relating to the intercalation mechanism for a wide variety of LiC$_x$ compounds. In the course of this process, we believe to have shown that Density Functional Tight Binding (DFTB) is a superior approach for modelling intercalation compared with force field methods, including the more sophisticated machine learning approaches (\emph{e.g.} the GAP by~\cite{Fujikake2018} requires a manual correction term for lithium-lithium interactions which our method does not). Furthermore, we share key details and choices along this process and thus provide guidance for similar endeavours in the future.

\begin{acknowledgement}

CP gratefully acknowledges funding from the German Research Foundation (DFG -- Deutsche Forschungsgemeinschaft) through grant~\#~PA~2932/1-1.
\end{acknowledgement}

\begin{suppinfo}
The Supporting Information is available free of charge at ...

Supplementary discussion, including computational details, PSO cost function, choice of basis set, {\tt GPrep} hyperparameters, description of training sets, full validation.
\end{suppinfo}

{\footnotesize
\bibliography{references}}

\end{document}


\section{Computational details}

DFT reference calculations have been performed by means of the all-electron electronic structure code {\tt FHI-aims}~\cite{Blum2009, Havu2009, Ren2012} with \emph{light} settings and default \emph{tier-2} basis sets. We used the GGA--based PBE exchange correlation functional~\cite{Perdew1992, Perdew1996} with the atomic Zero-Order Regular Approximation (ZORA) relativistic correction~\cite{Blum2009}, a converged $k$-point grid of $7\times 7\times 7$ and default convergence criteria for the self-consistent cycle. DFTB calculations have been executed via the {\tt DFTB+} code~\cite{Aradi2007} version 19.1, with a $k$-grid matching that of the DFT calculations. The convergence criterion for the self-consistent charges has been set to a relative difference of $10^{-8}$. To facilitate convergence, we used the standard Fermi distribution filling with an electron temperature of $0.0001$ K and the Anderson mixer~\cite{Eyert1996} with a mixing parameter of $0.1$. Geometries have been constructed and analysed by means of the Atomic Simulation Environment ({\tt ASE})~\cite{Bahn2002} which we also used as a basis framework for all force- and energy-calculations, structure relaxations (specifically using the BFGS algorithm~\cite{Shanno1985}) and barrier calculations. For the latter, we employed the Climbing Image Nudged Elastic Band (CI-NEB)~\cite{Henkelman2000, Henkelman2000a} algorithm, with transition paths consisting of five images unless stated otherwise and a FIRE optimizer~\cite{Bitzek2006}. In those instances when we employed dispersion correction, we used the Lennard Jones (LJ) approach~\cite{Zhechkov2005} with Universal Force Field (UFF) parameters~\cite{Rappe1992} for DFTB and range-separated self-consistent screening Many Body Dispersion (MBD@rsSCS, throughout the text referred to as MBD)~\cite{Tkatchenko2012, Ambrosetti2014} for DFT, with default range separation parameters. 
The PSO-based parametrization routines used for the electronic part of DFTB were implemented within our own development version (available upon request) of the {\tt hotbit} framework (\href{https://github.com/pekkosk/hotbit}{https://github.com/pekkosk/hotbit}),~\cite{Koskinen2009} augmented, among other features, with the Woods-Saxon confinement potential form, cost function modules for selected properties, and an interface with the PSO python package {\tt pyswarm} (\href{https://pythonhosted.org/pyswarm/}{https://pythonhosted.org/pyswarm/}). The GPR-based DFTB repulsive potential was generated with the python package {\tt GPrep}~\cite{Panosetti2020}, available on Zenodo at \href{https://zenodo.org/record/3697913}{https://zenodo.org/record/3697913}. As for all the systems of interest two or more lithium atoms remain separated by large distance, we hereby parametrize the repulsive potentials for C-C and Li-C only. A Li-Li repulsive potential is subject of ongoing work and will be released in the future.

\section{PSO cost function}

A band structure with $n$ bands, sampled at $m$ special points (or, more generally, at any set of $m$ k-points) may be represented by an array $BS$ of the following form:

\begin{equation}
 BS = (x_{00}, x_{01}, ..., x_{0n}, x_{10}, ..., x_{1n}, ..., x_{m0}, ..., x_{mn})
\end{equation}

where $x_{ij}$ represents the $j^{th}$ band energy at the $i^{th}$ special point, {\em e.g.}, $x_{00} = (\Gamma, E_{\Gamma, 0})$.

With such a representation one may map the band structure to a (symmetric) distance matrix D, of the following form:

\begin{equation}
 D = 
 \begin{pmatrix}
  0 & d^{01}_{00} & \cdots & d^{mn}_{00} \\
  d^{00}_{01} & 0 & \cdots & d^{mn}_{01} \\
  \vdots  & \vdots  & \ddots & \vdots  \\
  d^{00}_{mn} & d^{01}_{mn} & \cdots & 0 
 \end{pmatrix}
\end{equation}

The DFTB band structure will, by construction of DFTB, only include points belonging to the valence band and to the conduction band up to the basis set limit. Correspondingly, the DFT band structure representation for the PSO cost function is only built with points also accessible to DFTB. 

The special points and the number of compared bands have been selected empirically, with the goal of representing the most characteristic regions of the band structures properly (5 bands at the $K$ and the $\Gamma$ points for graphite, as well as the $W$ and the $\Gamma$ points for diamond, and 4 bands at the $H$ and the $\Gamma$ points for lithium).

The cost function for PSO is then chosen as the RMSD between the DFT and DFTB band structure distance matrices. 

The final electronic parameters were selected from a number of PSO runs with varying number of particles, where some runs included band structures for both elements in the cost function, while others were run separately for Li and for C. The two procedures did not yield significantly different parameters. The confinement constants were optimized in a parameter space varying in the following ranges:

\begin{center}
\begin{tabular}{l | c c }
 \hline
 Species    &   $r_0$       &   $r_{cut}$   \\
 \hline
 C          &   1-3 Bohr    &   3-7 Bohr    \\
 Li         &   2-5 Bohr    &   4-8 Bohr    \\
 \hline
\end{tabular}
\end{center}

The carbon confinement was subsequently adjusted by hand to be slightly looser in order to improve the descriptions of CC bond lengths. 

The onsite energies $\epsilon$ were only slightly optimized starting from initial guesses taken from Ref.~\cite{Wahiduzzaman2013}. The Hubbard U parameters were kept fixed at those taken from the same reference.

The optimized electronic parameters are the following:

\begin{center}
\begin{tabular}{l | c c c c}
 \hline
 Species & $\epsilon_s$ & $\epsilon_p$ & $r_0$ & $r_{cut}$ \\
 \hline
 C      & -0.5053 Hartree & -0.1942 Hartree & 3.11 Bohr = 1.64 \AA{} & 6.42 Bohr = 3.40 \AA{} \\
 Li     & -0.1057 Hartree & -0.0087 Hartree & 6.42 Bohr = 3.40 \AA{} & 6.72 Bohr = 3.55 \AA{} \\
 \hline
\end{tabular}
\end{center}

\section{Lithium $\{sp\}$ vs $\{s\}$ basis}

As mentioned in the main text, the electronic parameters for lithium were optimized in an $\{sp\}$ basis, while only the $s$ orbital was included in the construction of the final Slater Koster tables. The choice is motivated in the following.

In an initial tentative GPR parametrization using force residues calculated with an $\{sp\}$ DFTB basis, we noticed that the GPR model struggled to produce satisfactory results. In particular, we noticed a consistent trend to produce systematically too short interlayer spacings regardless of the choice of the training set, as well as systematically overestimated forces, and, as a consequence of both the above, systematically too large diffusion barriers.

\begin{figure*}[t]
\centering
\includegraphics[width=0.95\linewidth]{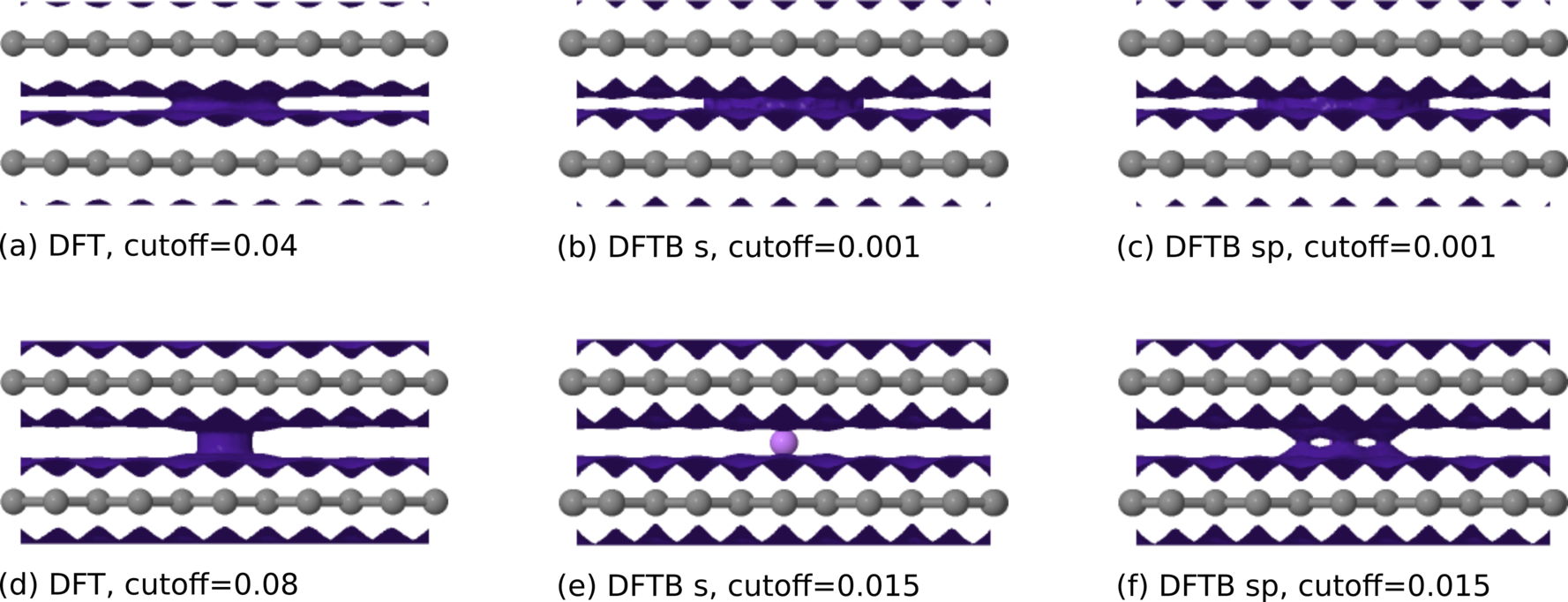}
\caption{Isosurface plots of total charge density with DFT (left), DFTB in a Li $s$ basis (middle) and DFTB in a Li $\{sp\}$.}
\label{fig:cubes_total}
\end{figure*}

Intuitively, this trend can be attributed to---probably pathological---overbinding. Substantial support of this hypothesis comes from a qualitative analysis of the DFT and DFTB charge densities. Figure~\ref{fig:cubes_total} shows total self-consistent charge densities for DFT (left), DFTB with an $\{s\}$ lithium basis (middle), and DFTB with an $\{sp\}$ lithium basis (right). The isosurface cutoff for the top row was chosen such that the valence $s$ density for lithium is showing similarly for all three. Hereby, both DFTB $\{sp\}$ and $s$ exhibit nearly identical features. In the bottom row, a smaller cutoff is selected such that the contribution from the more contracted $p$ density is revealed (for DFT, the Li core density is visible, which is not present in DFTB). The additional spurious bonding charachter for DFTB in an $\{sp\}$ is clearly visible in panel (f).

\begin{figure*}[t]
\centering
\includegraphics[width=0.95\linewidth]{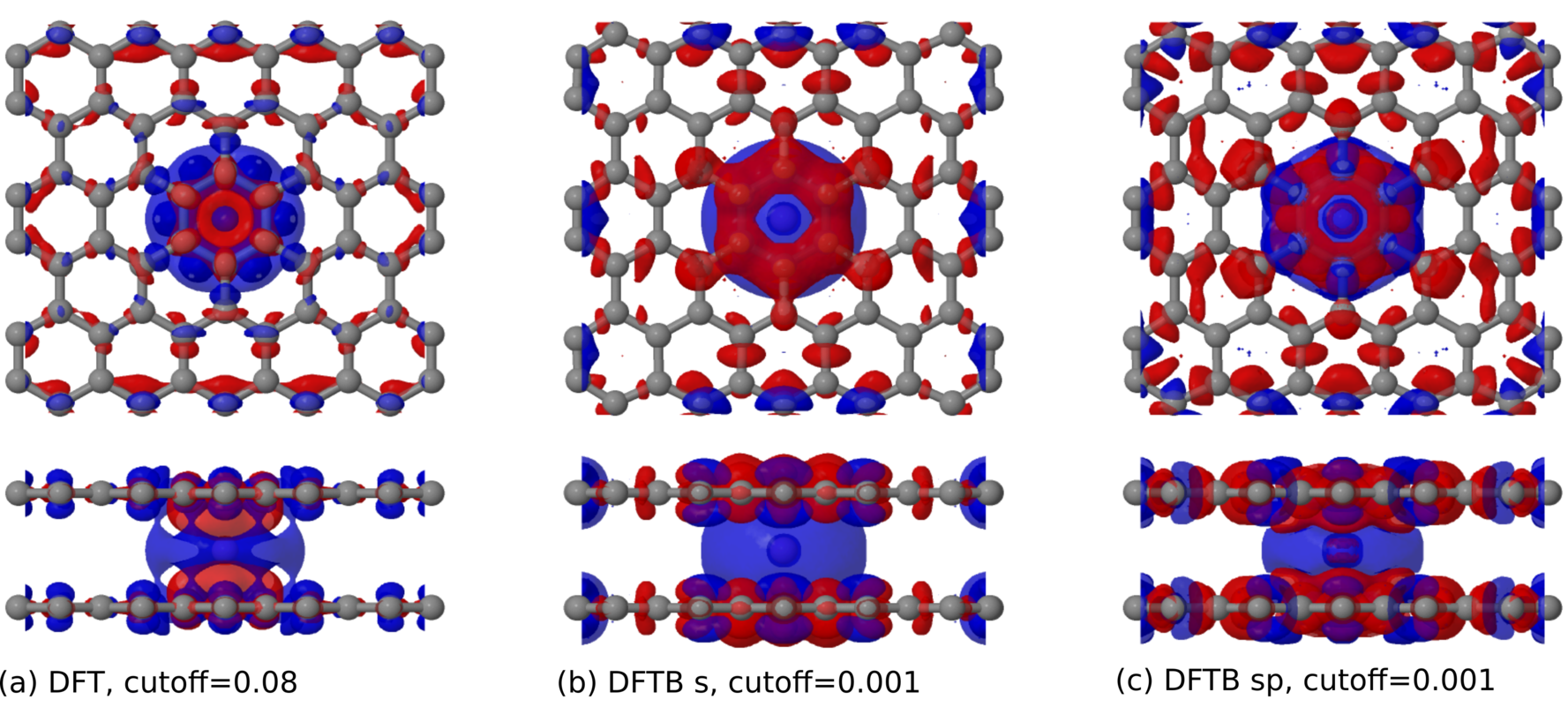}
\caption{Isosurface plots of charge density difference between the interacting graphite+lithium system and the sum of non-interacting fragments, with DFT (left), DFTB in a Li $\{sp\}$ basis (middle) and DFTB in a Li $\{sp\}$. Red represents charge accumulation, blue represents charge depletion.}
\label{fig:cubes_diffs}
\end{figure*}

The same effect is visible when analysing the charge density difference between the interacting system and the sum of fragments, as a qualitative measure of charge transfer between the graphite host and the lithium. This is shown in Figure~\ref{fig:cubes_diffs} . Of note, the densities shown here were calculated on a non-periodic slab, therefore the finite-size edge effects are to be ignored: we shall only focus on the central region depicting the interaction between the lithium and the graphitic host.

Despite the fact that (as shown in the top view), the charge transfer description {\em across} a single graphite layer is better captured in the $\{sp\}$ basis, the pitfall of the latter is evident when inspecting the charge transfer {\em between} graphite layers and the lithium atom (side view).

DFTB in an $\{sp\}$ basis here shows high extra electron density (red crown-shaped feature in panel (c)) which in the DFT is much more localized in the center of the benzene ring above the Li and, in the vicinity of the Li, masked by a depleted area. In the $\{sp\}$ basis the electron density is not sufficiently transferred to the graphene layers, which manifests as partial covalent bonding---in particular, once we start moving the Li across a barrier to the next ring.

Additional arguments in favour of an $s$-only (albeit $sp$-optimized) basis, at least for the time being, are given in Section~\ref{sec:fm}, based on force-matching between DFT forces and DFTB forces including the {\tt GPrep} repulsion. However, those are repulsion-dependent and therefore any discussion is limited to the parametrizations (with $V_{rep}$) we can access with the particular dataset we produced throughout this work, while the arguments above are pertinent to the electronic part only, and therefore more general.

As a final remark, we stress that DFTB operates in a truly minimal (valence only) basis set. While it is generally true that larger basis sets will produce more accurate electronic properties, this is not necessarily valid when a basis set is so heavily truncated. With such a limited budget, the addition of a single basis function may introduce distortions that occasionally end up, {\em de facto}, worsening the description of the electronic structure.

\begin{figure*}[t]
\centering
\includegraphics[width=0.45\linewidth]{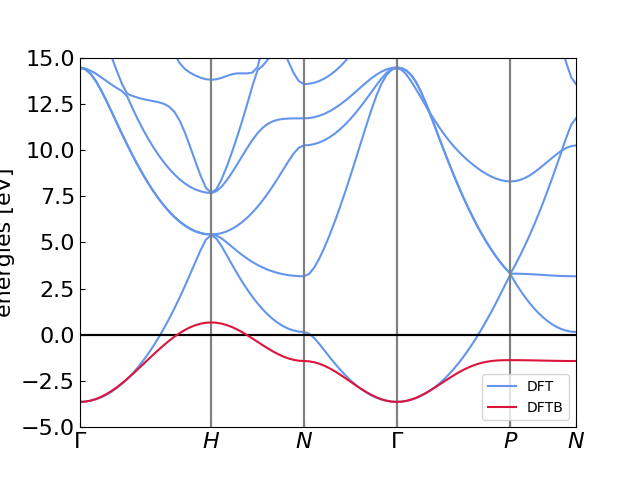}
\includegraphics[width=0.45\linewidth]{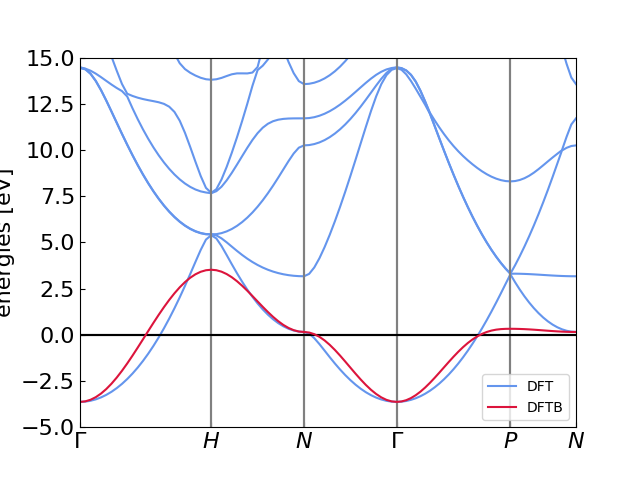}
\caption{Lithium band structures calculated with DFTB, with electronic parameters obtained optimizing the curvature at the gamma point (left) and the overall valence band (right).}
\label{fig:bands_s_only}
\end{figure*}

On the other hand, optimizing the lithium $2s$ orbital only in the PSO---not surprisingly---produces truly unconvincing band structures regardless of the choice of the technicalities of the cost function ({\em e.g.} either targeting the best curvature at the gamma point, or the best overall valence band), as shown in Figure~\ref{fig:bands_s_only}. The latter option has a decent overall dispersion, at least in terms of energy range, but clearly all the curvatures are wrong. The first option sacrifices the range of the dispersion in favour of a correct curvature at the gamma point. With confinement radii around 3 Bohr, both options result in a more compressed lithium atom w.r.t. the $\{sp\}$-optimized confinement, with the risk of producing systematically underbound structures due to reduced overlap. This may in principle be compensated by the repulsive contribution---which, we remind, is known to partially correct the imperfections of the electronic part. However, there is only so much extent to which the repulsive potential can actually do that effectively, while remaining acceptably smooth and transferable.

In the light of all the above, we prefer to rather work with the correct confinement for the complete valence basis---just ``switching off'' the $p$ contributions to the Hamiltonian and overlap integrals. This presents the additional advantage that the parameter set can be updated at will, in perspective, to include the full $\{sp\}$ interaction without having to reconsider the entire parametrization process, but rather focusing on the GPR part only. For future such revisions of the parameters, we make both Slater Koster tables and both {\tt GPrep} datasets available upon request, keeping in mind that, for a working $\{sp\}$ parametrization, the datasets most likely need to be extended.

\section{GPR repulsion hyperparameters}

Our implementation is essentially a simplified form of the algorithm developed by Cs\'anyi {\em et al.} for the generation of the already mentioned interatomic GAP potentials.~\cite{Bartk2010} Similarly, we here employ gaussian kernels of the form:

\begin{align}
k_{SE}(\vec{x_1}, \vec{x_2}) = \delta^2\exp\left(\frac{-(\vec{x_1} - \vec{x_2})^2}{2 \theta ^2}\right)  \label{GPR:eq:1}
\end{align}

\begin{figure*}[t]
\centering
\includegraphics[width=0.99\linewidth]{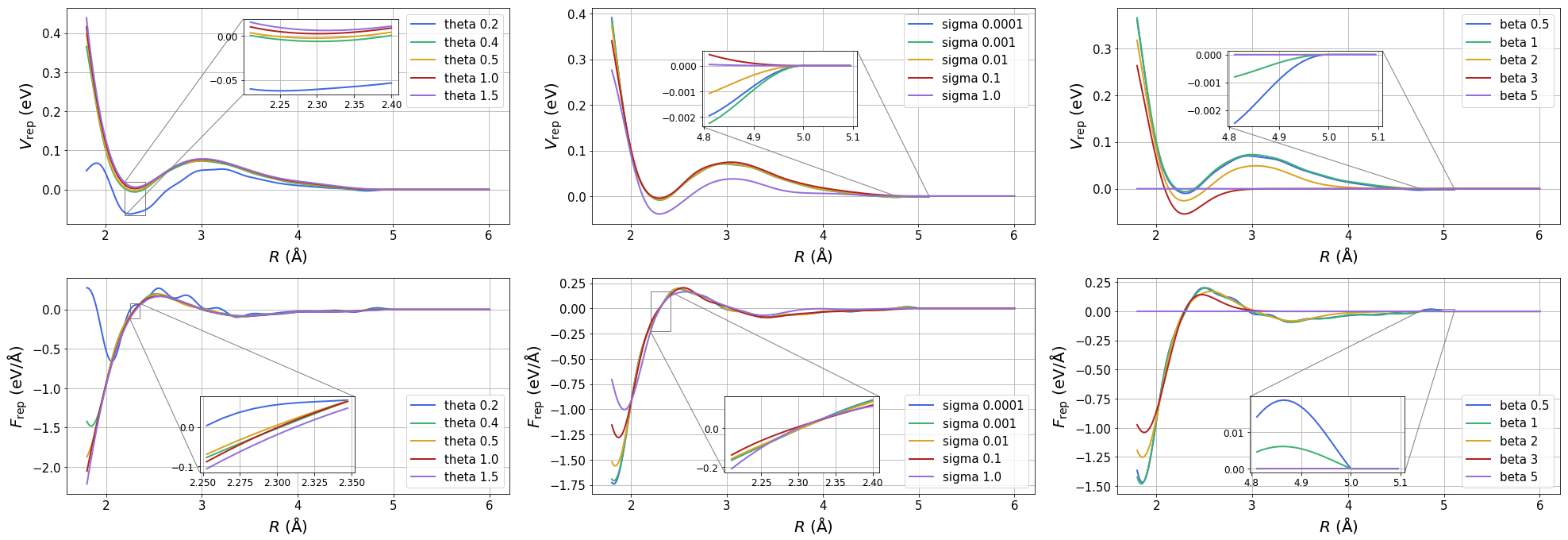}
\caption{Visualization of the impact which covariance parameter $\theta$ (left), data noise parameter $\sigma$ (middle) and the damping parameter $\beta$ have on the shape of an exemplary repulsion potential (top) and repulsive force (bottom).}
\label{fig:screens}
\end{figure*}

The parameters of $k_{SE}$ are commonly referred to as hyperparameters and need to be optimized. $\theta$ is a length scale within which the target values are similar. $\delta^2$ is the target variance in the prior distribution. The data noise $\sigma_n$ is also considered to be a hyperparameter. In fact, the posterior mean only depends on the quotient $\sigma_n / \delta$. In short, it can be said that $\theta$ is a measure of smoothness. The higher the value, the more our repulsion potential tries to ``avoid'' high curvatures and frequent curvature changes, which can have quite a critical effect (see Figure~\ref{fig:screens}, left). Similarly, but less importantly (see Figure~\ref{fig:screens}, middle), $\sigma_n$ should be just large enough to prevent over-fitting, but not much larger, since the posterior distribution will then avoid high amplitude targets. Based on these considerations and taking into account our present system via some potential-screening, we employ $\theta = 0.4$ and $\sigma = 2.\cdot 10^{-2}$. To fulfill the requirement for interatomic potential to vanish at large distances, the kernel is modified with a multiplicative cutoff function: 

\begin{equation}
f_{cut}(R) = \begin{cases}
e^{-\beta R}        & R < R_{cut} - d\\
e^{-\beta R}g(R)    & R_{cut} - d < R < R_{cut}\\
0                   & R> R_{cut}
\end{cases}
\end{equation}

where
\begin{equation}
g(r) = \frac{\cos\left(\pi \frac{R-R_{cut}+d}{d}\right)+1}{2}\,.
\end{equation}

This damping function smoothly sets the potential to zero at $R_{cut}$ over an interval of width $d$. In practice, $e^{-\beta R}$ is nothing but the exponential prior of the GPR model, and also allows to put some extra exponential damping onto the tail. 
Overall, when selecting $\beta$, opposing effects need to be taken into account: too low a value will typically not smooth out the transition to zero enough, while too large a value may cause the loss of some (possibly) important physical subtleties. In light of these principles, we set $\beta = 1.0$, just large enough to ensure smoothness at the zero-transition, but not so large to completely flatten out the long-ranged potential tail (or even the entire potential, see Figure~\ref{fig:screens}, right).

\section{Training and validation sets}
\label{sec:sets}

\subsection{Set 1 - Rattled structures with varied unit cell volume}
As mentioned in the main text, we begin with a training set made of a balanced mix of Li$_n$C$_{36}$ super-cells ($n \in (0, 1, ..., 6)$, 20 structures each). With this, we loosely follow the procedure suggested in~\cite{Engelmann2018}. There, the author used a set of distorted diamond-structures to test the {\tt GPRep} framework. The distortion was done {\em i)} by scaling the unit cell volume by factors between 0.75 and 1.75 and {\em ii)} by randomly displacing (rattling) the atoms with a standard deviation of $20\%$ of the scaled equilibrium nearest-neighbour distance. The purpose of this set was to test performance and resilience of the model for highly asymmetric atomic environments, with forces of more than $300$~eV/\AA{}. In contrast, the goal in \emph{this} work is not to drive the {\tt GPRep} framework to its limits, but to achieve a parametrization that performs well for real systems which are not \emph{in} equilibrium, but close to it. Having said this, we still believe it is important to incorporate at least some strongly distorted structures into our training data which serve the purpose of guaranteeing a satisfactory level of transferability (a view also shared by other researchers working with parametrizations~\cite{Fujikake2018}). However, we would like to assure that these structures do not dominate the fitting procedure. Relatedly, we recognize, that quantitative accuracy in those high-force regions is not of great importance, as long as the qualitative behaviour there---ensuring that the system leaves these regions immediately---is intact. In light of these considerations we make the following adjustments:
\begin{itemize}
 \item We limit the factor for the volume scaling to only range from 0.94 to 1.5, thus not compressing stiff covalent bonds quite as much.
 \item Also, we modify the standard deviation of the rattling so that it does not amount to $20\%$ by default, but a random value between $0\%$ and $20\%$ instead, thus giving a higher emphasis to structures that are close to equilibrium.
 \item Finally, we geometrically pre-screen all our structures in terms of bond lengths, which we do not allow to be shorter than $75\%$ of the equilibrium bond lengths.
\end{itemize}

By these measures, we assure convergence of almost all (except 5) reference calculations and limit our appearing forces to below $70$~eV/\AA{}.

Set 1 therefore contains 135 structures and 15831 force components.

\subsection{Set 2 - Addition of transition states}

In Set 2, we added structures extracted from Li$_2$C$_{36}$ stage I and II NEBs (20 structures each) and from a LiC$_{36}$ NEB (45 structures), with gaussian rattling applied to the moving lithium atom only in the plane perpendicular to the diffusion path. Concomitantly, we removed the rattled and scaled Li$_n$C$_{36}$ structures with $n > 1$. 

Set 2 contains 120 structures and 13386 force components.

\subsection{Set 3 - Dispersion-corrected force residues}

In Set 3, we replaced the DFTB-DFT force residues from all the structures Li$_n$C$_{36}$ with $n \in (0, 1)$ (including NEB) with DFTB(LJ)-DFT(MBD) force residues. This amounts to about 70\% of structures and 66\% of force components. The choice of limiting the reference MBD forces to structures containing no lithium atoms or only one is motivated by the observation (also confirmed in Ref.~\cite{Gould2016}) that MBD (in the implementation with analytical forces) tends to exhibit numerical issues for structures containing more than one Li atom, due to limitations of the Hirshfeld partitioning. As those issues occasionally produce negative polarizabilities, we only picked those calculations for which this did not happen.

\subsection{Validation set}

The validation set includes 45 structures (for a total of 5265 force components) extracted from a LiC$_{36}$ NEB, with additional gaussian rattling (different from that applied in Set 2). We focused the training set on NEB as we intended to place emphasis on analysing parallel ({\em i.e.} in the $xy$ plane) and perpendicular forces separately, which are of particular importance in the diffusion processes which ultimately motivate the present work.

\section{Plateau plots}

In this Section, we report and discuss both $R_{cut}^{\mathrm{CC}}$ and $R_{cut}^{\mathrm{LiC}}$ scans of selected properties for all the training sets employed throughout this work.

For structural properties, we evaluate both C-C bond lengths and average interlayer distances for graphite, LiC$_{12}$ and LiC$_{6}$. As an additional benchmark, we evaluate the performance in prediction of forces. We choose the relative Root Mean Square Deviation (rRMSD) of forces and the Mean Absolute Force compared with the DFT references, as a measure of how much our predictions are scattered, and of whether they are scattered around the correct mean, respectively. Of note, a correct Mean Absolute Force does not necessarily correspond to correct forces overall, especially if---as in our case---the validation set contains approximately symmetric forces. This is also evident in the force matching plots in Section~\ref{sec:fm}.

To define whether or not points belong to a plateau, and whether or not we hit an adequate target of accuracy, we define the following thresholds (length units are in \AA{} and force units in eV/\AA{}):
\begin{center}
\begin{tabular}{l | c c c c c}
 \hline
 Quantity  & C-C bond length & Avg. layer spacing  & Mean Absolute Force  & Force rRMSD \\
 \hline
 Plateau if & \vtop{\hbox{\strut pt within 0.005}\hbox{\strut wrt prev/next}} & \vtop{\hbox{\strut pt within 0.02}\hbox{\strut wrt prev/next}} & \vtop{\hbox{\strut pt within 0.015}\hbox{\strut wrt prev/next}} & \vtop{\hbox{\strut pt within 0.02}\hbox{\strut wrt prev/next}} \\
 green if& \vtop{\hbox{\strut pt within 0.01}\hbox{\strut wrt expt}} & \vtop{\hbox{\strut pt within 0.06}\hbox{\strut wrt expt}} & \vtop{\hbox{\strut pt within 0.025}\hbox{\strut wrt DFT}} & \vtop{\hbox{\strut pt $0>0.5$ (C)}\hbox{\strut pt $0>0.2$ (Li)}} \\
 \hline
\end{tabular}
\end{center}

\begin{figure*}[h]
\includegraphics[height=0.49\linewidth]{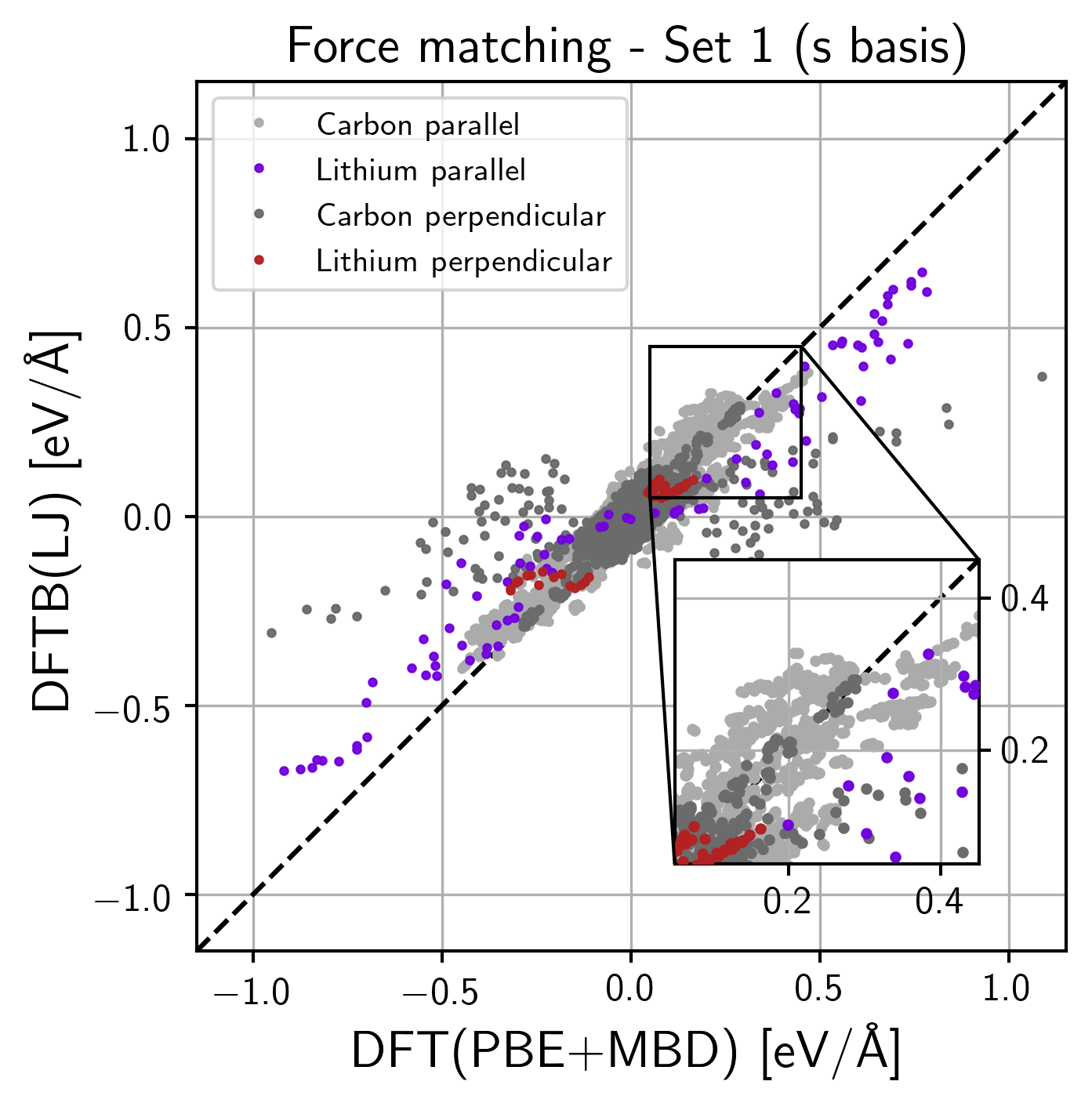}
\includegraphics[height=0.49\linewidth]{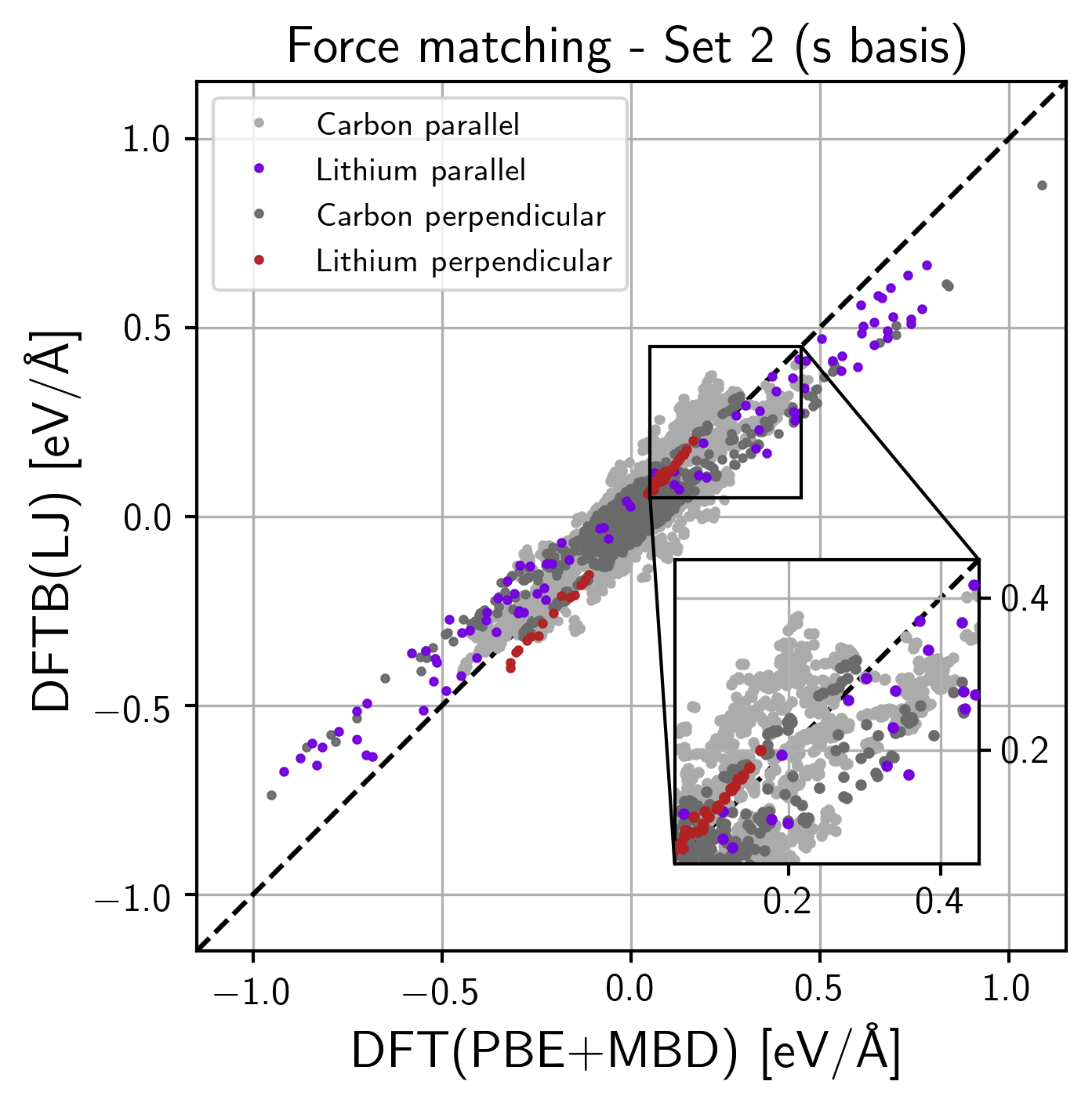}\\
\includegraphics[height=0.49\linewidth]{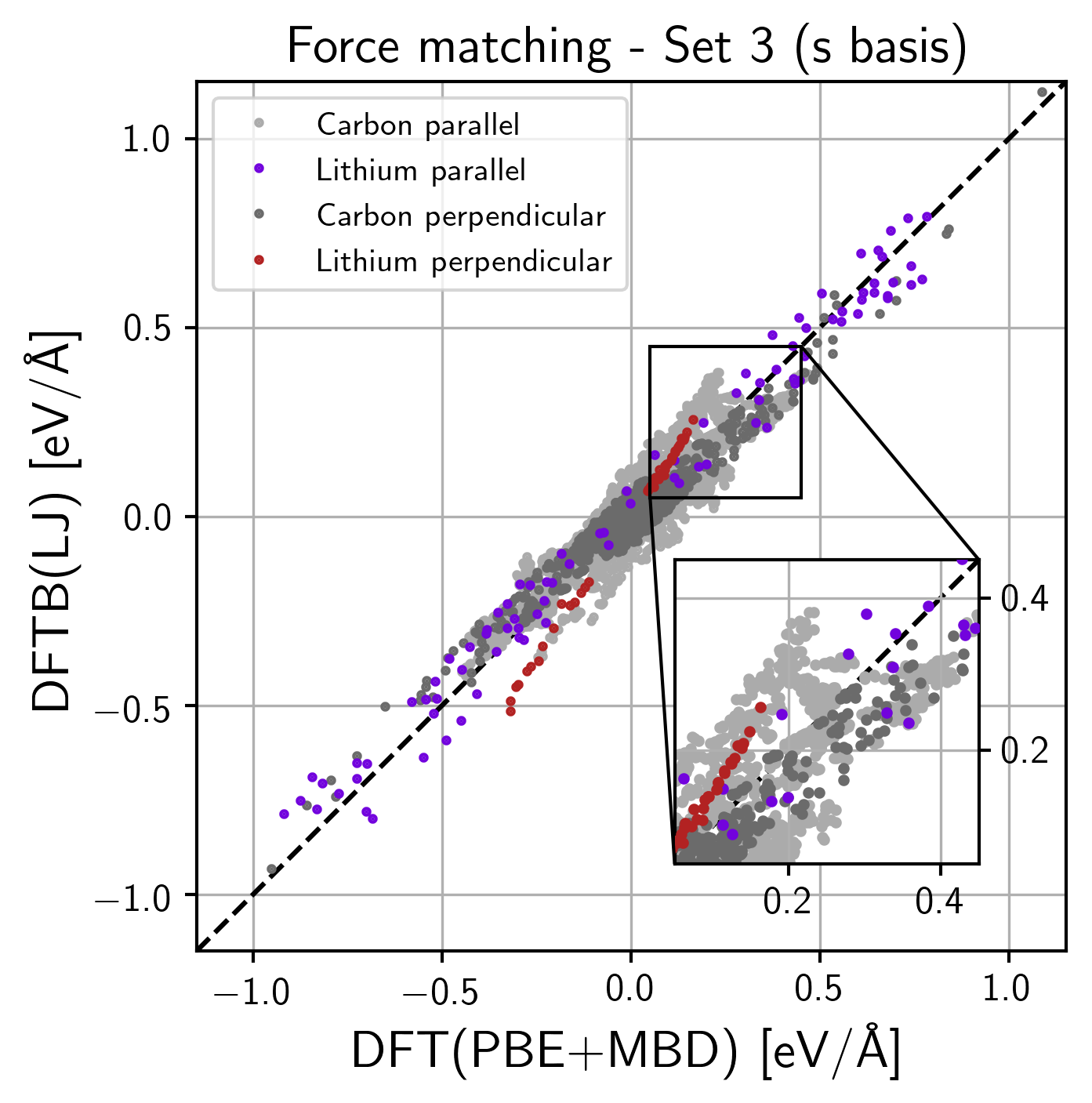}
\includegraphics[height=0.49\linewidth]{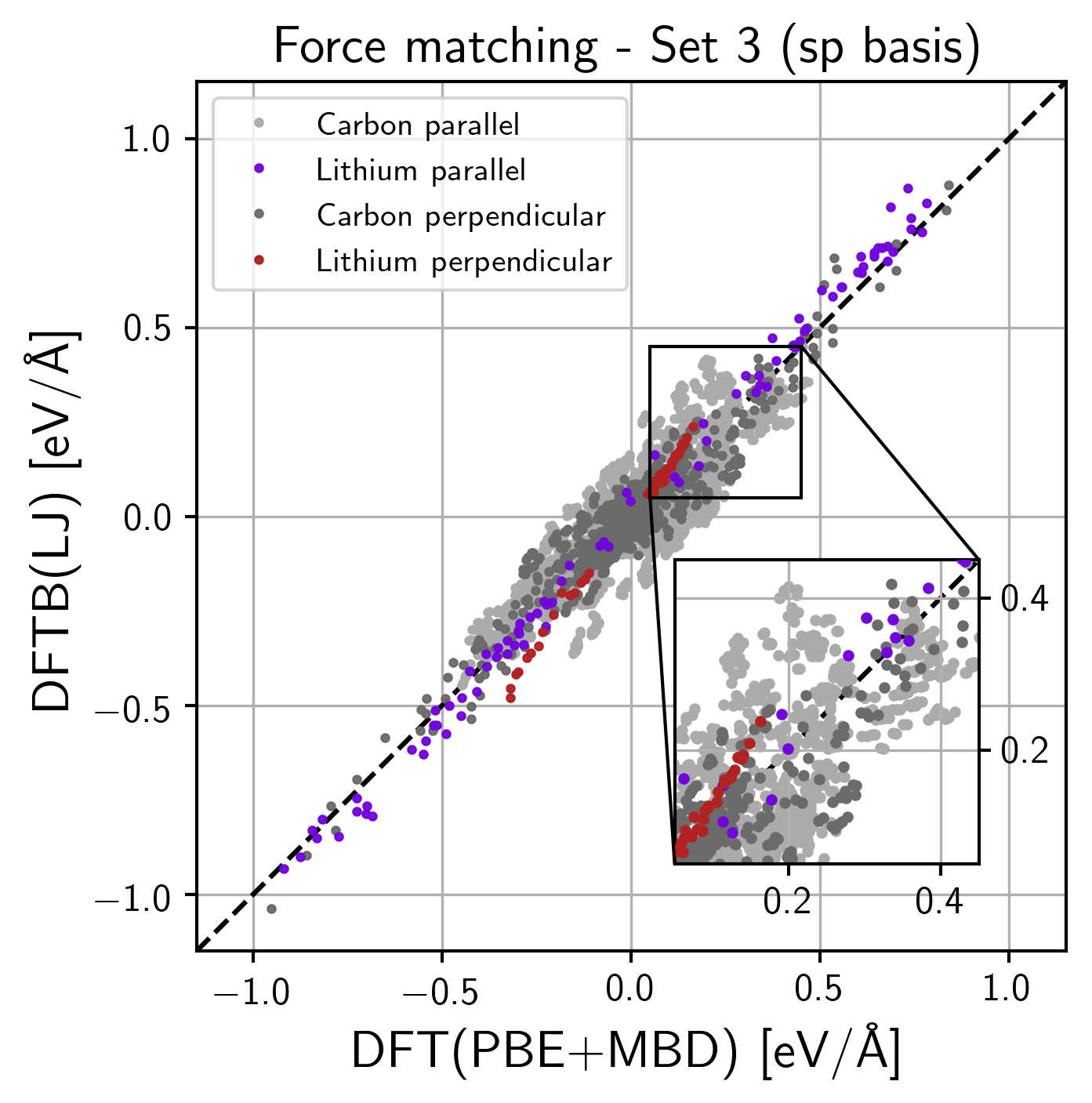}
\caption{Force matching plots for Sets 1 to 3, as well as Set 3 with an $\{sp\}$ basis, for the validation set defined in Section~\ref{sec:sets}.}
\label{Fig:force_matching_all}
\end{figure*}

Figure~\ref{Fig:force_matching_all} (top row) shows the C-C cutoff scan for a Li-C cutoff of 4.0~\AA{} for all the analysed quantities. The interlayer distances are correct for the entire range of C-C cutoff radii, and the C-C bond, in a smaller plateau, slightly underestimated (albeit with the correct ordering and within a still acceptable error). The Mean Absolute Force is only correct outside the plateau for the first two quantities, and the rRMSD of forces is beyond our chosen threshold for the entire C-C cutoff range. Arguably, although, the description of forces here appears relatively acceptable.

However, the second row shows how results for Set 1 are only ``accidentally'' correct for a Li-C cutoff of 4.0~\AA{}. After scanning the results for interlayer distance for the C-C cutoff (see main text), we investigate the Li-C cut-off in the same way. The Li-C repulsion potential does not influence our prediction of graphite in any way, so here the only relevant benchmarks are the LiC$_{12}$ and LiC$_6$ layer distances. For the Li-C cut-off, we barely observe a sizeable range of values for which the resulting layer distances are constant (and especially not constant at the correct value). We \emph{are} able to get a parametrization that predicts the correct result, but \emph{only} for a cutoff radius in the very close proximity of $4.0$~\AA{}. Additionally, the Mean Absolute Force and rRMSD of forces are largely off for every choice of Li-C cutoff, with the only acceptable values, again for $4.0$~\AA{} and not belonging to any plateau.

The two middle rows show the effect of the modifications introduced in Set 2. As we now observe Li-C plateaus for Li-C cutoff values larger than 4.0~\AA{}, from now on we show C-C scans for a Li-C cutoff of 5.0~\AA{}---the one we ultimately chose. The removal of higher saturated rattled and scaled compounds brings the C-C bond lengths to the desired level of accuracy, in the same C-C cutoff plateau discussed in the main text and in a larger Li-C plateau (not surprisingly, as the Li-C cutoff only influences this quantity indirectly, via minor modifications of the C-C $V_{rep}$ in the simultaneous fitting of both potentials in the {\tt GPrep} procedure). However, all the other quantities are systematically wrong for any choice of C-C cutoff. With respect to the Li-C cutoff, the modifications in Set 2 stabilize, to varying extents, all the targeted quantities, however not to the desired level of accuracy.

The two bottom rows show the same C-C and Li-C cutoff scans for Set 3. The introduction of MBD-corrected force residues finally yields stable {\em} and correct results for the largest range of both C-C and Li-C cutoff choices.

\begin{figure*}[h]
\centering
\includegraphics[height=0.93\textheight]{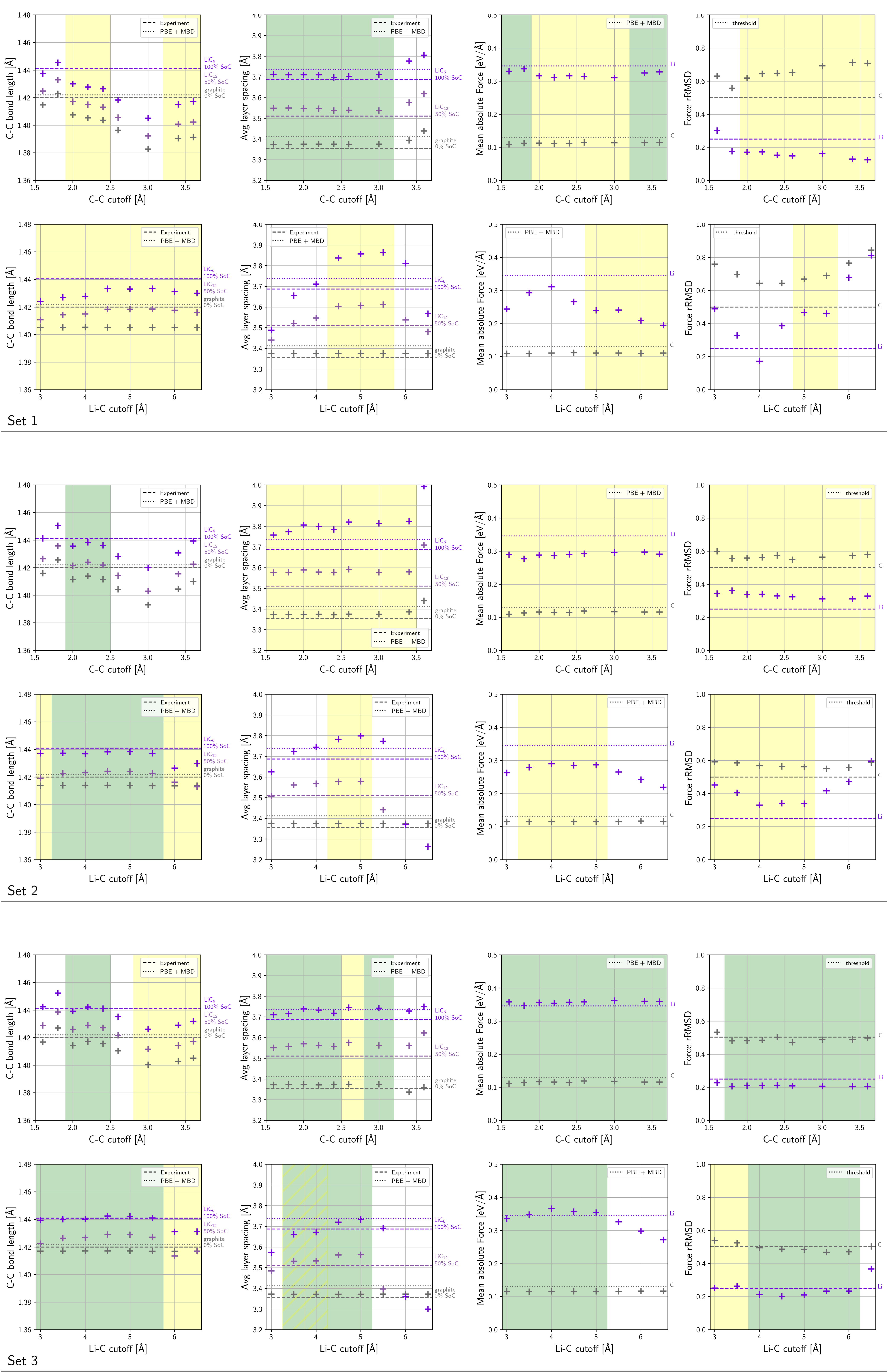}
\caption{C-C and Li-C repulsion cutoff scans for Set 1 (top), Set 2 (middle) and Set 3 (bottom). Evaluated properties are, left to right: C-C bond lengths, interlayer spacings, mean absolute force and rRMSD of forces.}
\label{Fig:plateaus_all}
\end{figure*}
\FloatBarrier

\section{Force matching}
\label{sec:fm}

\FloatBarrier

Training on Set 1 produces heavily scattered and severely underestimated carbon perpendicular forces. Lithium perpendicular forces are also underestimated, in line with the fact that the training set is lacking NEB geometries.

The addition of NEB geometries (Set 2) corrects the slope for lithium perpendicular forces, while the lithium parallel forces become less scattered, but underestimated. 

Set 3 presents overall the best compromise (with some residual carbon deviation), despite losing some accuracy in lithium perpendicular (not visible in the plateau plots which are dominated by parallel forces). The latter fact seems to actually be a fortunate accident, which might be responsible for correcting the layer spacings by placing some extra attraction in the $z$ direction. 

As a final remark, we compare the results for Set 3 in a Li $\{sp\}$ basis. First and foremost, a direct comparison with the same training set in a Li $\{s\}$ basis shows that the latter is visibly better overall, especially in the lower-force region. In addition to that, two main features here suggest that, at least with the current dataset, we cannot achieve a truly satisfactory parametrization in the Li $\{sp\}$ basis: {\em i)} there is a considerably larger scatter of carbon forces than all the training sets in an $\{s\}$ basis. While, generally, wrong slopes in force matching scatter plots indicate systematic errors that can be fixed by a different choice of training sets, a large scatter around the correct slope usually indicates that the model has hit its limit; and {\em ii)} much more subtly, some of the (both parallel and perpendicular) carbon forces appear ``split'' around two different slopes, a trend which we also observed, even more pronounced, with different tentative training sets not shown here. To be fair, this also happened for Set 1 in the $\{s\}$ basis and was corrected by a more careful choice of the training set. However, here our best training set does not eliminate the problem completely. We speculate that this residual structure in the scattered forces may be due to the fact that the addition of the Li $p$ orbital essentially breaks the hexagonal symmetry. Both the above suggest that the lower isotropy of the lithium orbitals in an $\{sp\}$ has an effect on the chemical environment, ``transferring'' error on the carbon forces. Given the quality of the results with the Set 3, Li $\{s\}$ basis parametrization, and given that a Li $\{sp\}$ produces too short interlayer spacings anyway (3.547~\AA{} for LiC$_6$ with Set 3), with dramatic effect on diffusion barriers, we do not feel the need to pursue the $\{sp\}$ parametrization any further in this work, but we once agan remark the possibility of doing so in future revisions.

{\footnotesize
\bibliography{references}}